\documentclass{article}
\usepackage{emulateapj}
\usepackage{epsfig}
\usepackage{graphicx}
\usepackage{color} 
\usepackage{natbib}
\newcommand{\ic}{IC\,443} 
\newcommand{\cxoulong}{CXOU~J061705.3$+$222127}
\newcommand{\ploc}{06$^{\rm h}$17$^{\rm m}$5.200$^{\rm s}$ $+$22\arcdeg21\arcmin27.52\arcsec}
\newcommand{\cxou}{J0617}
\newcommand{\ddd}{$d_{1.5}$}
\newcommand{\msun}{M$_{\odot}$}
\newcommand{\ergl}{erg~s$^{-1}$}
\newcommand{\kms}{km~s$^{-1}$}
\newcommand{\nh}{$n_{\rm H}$}

\newcommand{\ergcms}{erg~cm$^{-2}$~s$^{-1}$}
\newcommand{\ha}{H$\alpha$}
\newcommand{\oiii}{[O{\sc iii}]}

\newcommand{\cxo}{{\sl Chandra}}

\slugcomment{Accepted to Astrophysical Journal}

\begin{document}

\title{High Spatial Resolution X-Ray Spectroscopy of the \ic\ Pulsar Wind Nebula and Environs}

\author{
Douglas A. Swartz\altaffilmark{1},
George G. Pavlov\altaffilmark{2},
Tracy Clarke\altaffilmark{3},
Gabriela Castelletti\altaffilmark{4},
Vyacheslav E. Zavlin\altaffilmark{1},
Niccol\`{o} Bucciantini\altaffilmark{5},
Margarita Karovska\altaffilmark{6},
Alexander~J.~van~der~Horst\altaffilmark{7},
Mihoko~Yukita\altaffilmark{8,9},
Martin~C.~Weisskopf\altaffilmark{10}
}

\altaffiltext{1}
{USRA, Astrophysics Office, NASA Marshall Space Flight Center, ZP12, Huntsville, AL 35812, USA}
\altaffiltext{2}
{Department of Astronomy \& Astrophysics, Pennsylvania State University, 525 Davey Lab, University Park, PA 16802, USA}
\altaffiltext{3}
{Remote Sensing Division, Code 7213, Naval Research Laboratory, 4555 Overlook Avenue, SW, Washington, DC, USA}
\altaffiltext{4}
{Instituto de Astronomía y Física del Espacio (IAFE, CONICET-UBA), CC67, Suc.28, 1428, Buenos Aires, Argentina}
\altaffiltext{5}
{INAF - Osservatorio Astrofisico di Arcetri, L.go E. Fermi 5, I-50125 Firenze, Italy; INFN - Sezione di Firenze, Via G. Sansone 1, I-50019 Sesto F.no (Firenze), Italy}
\altaffiltext{6}
{Smithsonian Astrophysical Observatory, MS 4, 60 Garden Street, Cambridge, MA 02138, USA} 
\altaffiltext{7}
{Department of Physics, The George Washington University, 725 21 Street NW, Washington, DC 20052, USA}
\altaffiltext{8}
{The Johns Hopkins University, Homewood Campus, Baltimore, MD 21218, USA}
\altaffiltext{9}
{NASA Goddard Space Flight Center, Code 662, Greenbelt, MD 20771, USA}
\altaffiltext{10}
{Astrophysics Office, NASA Marshall Space Flight Center, ZP12, Huntsville, AL 35812, USA}

\begin{abstract}
Deep \cxo\ ACIS observations of the region around the putative pulsar, \cxoulong, 
in the supernova remnant IC443 reveal an $\sim$5\arcsec-radius ring-like
structure surrounding the pulsar and a jet-like feature oriented roughly
north-south across the ring and through the pulsar's location at \ploc\ (J2000.0 coordinates).
The observations further confirm that (1) the spectrum and flux of the central
object are consistent with a rotation-powered pulsar, 
(2) the non-thermal
spectrum and morphology of the surrounding nebula are consistent with a
pulsar wind and, (3) the spectrum at greater distances is consistent with
thermal emission from the supernova remnant. The cometary shape of the
nebula, suggesting motion towards the southwest, appears to be subsonic:
There is no evidence either spectrally or morphologically 
for a bow shock or contact discontinuity; the nearly circular ring 
is not distorted by motion through the ambient medium;
and the shape near the apex of the nebula is narrow. 
Comparing this observation with previous observations of
the same target, we set a 99\% confidence upper limit to the proper motion
of \cxoulong\ to be less than 44~mas~yr$^{-1}$ (310~\kms\ for a distance of 1.5 kpc), 
with the best-fit (but not statistically significant) projected direction toward the west.

\end{abstract}


\section{Introduction}

Pulsars lose their rotational energy in the form of magnetized relativistic winds.
Being confined by their environment, these winds produce luminous pulsar wind 
 nebulae (PWNe) with shapes and spectra determined by the pulsar emission geometry,
 magnetic field, and particle energy distribution as well as by the properties of the 
 ambient medium and the motion of the pulsar relative to its surroundings
 \citep[e.g.,][]{KarPav08}.
This is particularly true of pulsars still within their natal supernova 
 remnant (SNR).
PWNe evolve as they
 first expand into unshocked SN ejecta, then interact with the SNR reverse shock,
 the hot SN debris, and finally the surrounding interstellar medium
 \citep[see, e.g., the review by][]{GaeSla06}.
Moreover, the pulsar may obtain a high space velocity at birth, move
 supersonically through its surroundings, and produce a cometary PWN 
 with a bow shock that may be visible in X-rays and optical emission lines
 \citep{Buc02a}.

A particularly intriguing example of the PWN phenomena
 is \cxoulong\ (also known as G189.22$+$2.90 and as 1SAX\,J0617.1$+$2221; hereafter \cxou)
 located at the southern edge of the SNR shell \ic\ where the shell
 interacts with a molecular cloud \citep[e.g.,][]{Lee08}. 
This alignment suggests this PWN is entering a rare transitional stage where
 the overall morphology of the PWN is affected by its interaction
 with the SNR shell.
While it shows a distinct cometary structure suggesting supersonic motion
 \citep{Olb01,BocByk01,Gae06},
 the PWN is oriented $\sim$50\arcdeg\ away from the direction expected
 if the pulsar originated at the center of the \ic\ remnant.
This casts some doubt as to the physical association of this object and \ic\
 as pointed out by \citet{Gae06}.

A hard, non-thermal, X-ray source located in the southern portion of 
 \ic\ was first discovered by \citet{Keo97} with {\sl ASCA}.
The source was later observed with {\sl BeppoSAX} 
 \citep{BocByk00}, with {\sl XMM-Newton} \citep{BocByk01},
 and with \cxo\ ~for $\sim$10~ks using the 
 Advanced CCD Imaging Spectrometer (ACIS) I-array \citep{Olb01} 
 and again for $\sim$37.5~ks using the S-array \citep{Gae06}.

The latter three observations had sufficient angular resolution to 
 study the pulsar\footnote{A pulsar is inferred by the nature of its 
 X-ray properties although no pulsations have yet been observed from the compact object.
 } 
 and many details of the PWN and surroundings.
All three studies noted the soft thermal spectrum of the central point source
 is consistent with emission from the surface of a neutron star.
\citet{Olb01} first identified the comet-shaped morphology of the nebula
 and argued it could be explained by supersonic motion causing a pulsar wind 
 to terminate in a bow shock and flow downstream in a synchrotron tail.
The deeper observation obtained by \citet{Gae06} revealed a surface brightness plateau 
 extending to $\sim$25\arcsec\ downstream 
 of \cxou\ which they interpreted as the backward termination shock.
They used the locations of the forward and backward shocks to deduce a low Mach number, $\sim$1.2,
 for the flow and argued this was indicative of motion through relatively hot (shocked)
 supernova debris.
\citet{Gae06}
 also identified two compact knots of emission a few arcseconds north and south of \cxou.
\citet{BocByk01} showed that the 
 hard power-law spectrum at the core of the nebula softens with distance 
 from \cxou\ and is consistent with synchrotron cooling models.

Here, analysis of a much deeper, $\sim$152~ks, ACIS S-array
 observation is presented.
An overview of the observation and data reduction methods is 
 given in Section~\ref{s:dataprep}
 followed by analysis of the X-ray morphology, a comparison of the
 X-ray to 
 optical emission line images, and analysis of 
 the X-ray (and radio) spectrum at high spatial resolution (Section~\ref{s:results}). 
The bright source, \cxou, is confirmed to display the physical 
 properties of a pulsar (though no pulsations are detected) 
 that powers a bright, cometary-like, roughly 2\arcmin$\times$1.5\arcmin\ PWN. 
Within the nebula is a nearly circular, $\sim$5\arcsec\ radius, 
 ring of enhanced X-ray emission enclosing a jet-like feature oriented roughly north-south
 and passing through the position of \cxou.
The results are then interpreted (Section~\ref{s:discussion}) in terms of synchrotron 
 emission from relativistic particles generated by the pulsar and confined
 by the surrounding medium.
 
A distance of $d=1.5$~kpc (1\arcmin$=$0.42~pc) to \cxou\ is adopted
 following \citet{WelSal03} but note, as pointed
 out above, there remains some doubt as to the physical
 association of \cxou\ and the SNR \ic. Therefore, distances are given 
 in units of $d_{1.5}=d/1.5$~kpc where appropriate.

\section{Data \& Methods} \label{s:dataprep}

\subsection{X-ray Observations} \label{s:dataxray}

\cxou\ was observed with the \cxo\ X-ray Observatory
 Advanced CCD Imaging Spectrometer on 2012 February 6 (ObsID 13736)
 for  $\sim$107.6~ks and again
 on 2012 February 8 (ObsID 14385) for 44.5~ks.
Both observations used an identical instrument configuration
 which placed \cxou\ near the back-illuminated CCD (S3) aimpoint.
The observations were taken in full-frame timed exposure mode using 
 the standard 3.2~s frame time and the VFAINT telemetry format.
 
The data were reprocessed using the Chandra X-ray Center script 
 {\tt acis\_process\_events} (CIAO v. 4.5) to apply the latest 
 time-dependent gain correction (CalDB 4.5.5) and to flag potential 
 cosmic-ray background events.
Only events imaged on the back-illuminated CCD S3 are used
 in this work.
Point-like X-ray sources in the field were identified using the 
 source-finding utility in the {\tt lextrct} image-analysis 
 program  \citep{Ten06}.
The locations of sources identified in the two event lists were 
 compared and were found to be in sufficient agreement for most purposes so
 that the two event lists were simply combined using the FTOOLS utility {\tt fmerge}
 to produce a 152~ks duration final event list.
In addition, a sub-pixel image was created by rebinning the events 
 to $1/2$ of the nominal (0.492\arcsec) ACIS pixel size
 in order to examine the finer details of the 
 X-ray morphology in the immediate vicinity of \cxou\ (section~\ref{s:ring}).

Spectral fitting reported here uses the XSPEC, v.~11.3.2g, 
 spectral-fitting package \citep{Arn96}.  
Spectral redistribution matrices and ancillary response functions
 were generated using the CIAO tools {\tt mkacisrmf},  {\tt mkarf}
 and {\tt mkwarf}.
Spectral analysis is confined to events within
 the 0.5$-$8.0~keV energy range. 
The energy range spans only 0.5$-$5.0~keV 
 in those cases where a soft thermal-dominated spectrum produces a
 negligible number of high-energy source counts
 as is the case outside the PWN region.

The X-ray surface brightness and X-ray color or hardness ratio vary
 noticeably across the $\sim$8\arcmin$\times$8\arcmin\ S3 field of view.
The `contour binning' method of \citet{San06} was therefore applied to the merged 
 data set in order to map the physical properties, deduced from spectral
 model fitting, in the regions containing \cxou, its PWN, and 
 portions of the surrounding diffuse thermal plasma.
The contour binning method 
 defines spatial bins (for spectral analysis) which cover contiguous 
 regions of similar surface brightness and is motivated by the fact 
 that surface brightness variations often follow
 changes in physical properties, such as those at shock fronts, in the X-ray-emitting medium.
 
For this purpose, the identified point sources were first removed from
 an image constructed from the merged event list and back-filled 
 with a sampling of the adjacent background using the
 CIAO tools {\tt roi} and {\tt dmfilth} 
 following the diffuse emission science 
 thread.\footnote{ http://cxc.harvard.edu/ciao/threads/diffuse\_emission/}
The resulting image was then accumulatively smoothed using a signal-to-noise ratio 
 threshold of 30 and contour-binned using a signal-to-noise threshold of 100 using
 the {\tt accumulate\_smooth} and {\tt contbin} utilities, respectively,
 from \citet{San06}.
  
An earlier observation of \cxou\ (ObsID 05531), taken 2005 January 12
using a similar instrument configuration, is used in analysis of the proper motion of \cxou\
(section \ref{s:proper}).

\subsection{Optical Observations} \label{s:dataoptical}

The analysis presented here also makes use of 
 \ha\ and \oiii\ images obtained 23 Nov 2012
 using the Astronomical Research Cameras\footnote{http://www.astro-cam.com} E2V 230-42 CCD
  ($\sim$0.493\arcsec~pixel$^{-1}$) mounted on the 
 0.9~m Southeastern Association for Research in Astronomy telescope at Kitt Peak Observatory.
Seeing conditions were $\lesssim$2\arcsec.
Contemporaneous R and V images were used for continuum subtraction from \ha\ and \oiii,
 respectively.
The continuum-subtracted images were registered to the X-ray data using 2MASS objects
 common to both wavelengths.

\subsection{Radio Observations} \label{s:dataradio}

Archival observations of \ic\ at 330~MHz, 4.8~GHz, and 8.4~GHz,
 obtained with the Very Large Array (VLA),\footnote{The Very Large Array of the 
 National Radio Astronomy Observatory (NRAO) is a facility 
 of the National Science Foundation operated under cooperative agreement by 
 Associated Universities, Inc.} 
  were also analyzed and compared to the X-ray data.
All data were calibrated and reduced using the 
 Astronomical Image Processing System (AIPS) data reduction 
 package\footnote{http://www.aips.nrao.edu}. 

Multiple 330~MHz observations were carried out from 2005 to 2007 using several VLA configurations 
 \citep{Cas11} so that the final combined image is sensitive to a wide range of
 spatial structures from $\sim$6$^{\prime\prime}$ up to $\sim$70$^{\prime}$
 and therefore samples both the PWN and the surrounding extended \ic\ SNR.
See \citet{Cas11} for further details of the 330~MHz observations and data reduction.
The final combined image has an angular resolution of 17\arcsec\ and a 
 background noise level of 1.7~mJy~beam$^{-1}$ 
 after correcting for primary beam attenuations.
The average SNR contribution in the region surrounding \cxou\ is 15~mJy~beam$^{-1}$.

The 4.8~GHz observation was taken in the C configuration 
 on 1997 August 26 and the 8.4~GHz observation in D configuration
 on 1997 December 31.
The 4.8~GHz observation was taken with two 50~MHz tunings centered at 4.8851~GHz and 4.8351~GHz
 and the 8.4~GHz observation was taken in two 50~MHz bands centered at 8.4351~GHz and 8.4851~GHz.
Both observations used 3C~138 as the flux calibrator and 0632+103 as a phase calibrator.
These observations were first analyzed by \citet{Olb01}.
These data are only sensitive to structures on scales up to a 
 little over 2$^{\prime}$ and thus are not able to sample
 the emission over the entire nebula as is possible at 330~MHz.
Here, they were re-calibrated and reduced using 
 standard data reduction techniques applied within AIPS.
The data were self-calibrated in two rounds of phase-only calibration.
The final, full resolution, 4.8~GHz image rms is 0.18~mJy~beam$^{-1}$ 
 with a beam size of $\sim$5.5\arcsec$\times$ 3.8\arcsec\ and the
8.4~GHz image rms is 0.076~mJy~beam$^{-1}$ with a beam size of 
 $\sim$8.1\arcsec$\times$7.1\arcsec.

\section{Results} \label{s:results}

\subsection{General X-ray Morphology} \label{s:spatial}

The deep \cxo\ image (Figure~\ref{f:fullS3image}) 
 reveals \cxou\ as the brightest point-like
 object in the field surrounded by a high-surface-brightness nearly-circular 
 ring.
There are two `spokes' apparently connecting \cxou\ to the ring oriented
 roughly north-south which are possibly pulsar jets.
Exterior to the ring is a comet-shaped nebula with its major axis
 oriented $\sim$50\arcdeg\ East of North and with its apparent forward direction
 to the Southwest.
The cometary nebula has a hard X-ray spectrum as was noted 
by \citet{Olb01} 
 and \citet{Gae06}.
Predominantly soft X-ray emission surrounds the 
 roughly 2\arcmin$\times$1.5\arcmin\ cometary nebula
 and extends throughout the S3 field of view (along with a few faint unrelated point-like sources). 
The surface brightness of this extended emission is spatially non-uniform 
 varying by factors of 4 to 5 over spatial scales of order 1\arcmin.
As shown by spectral analysis (section~\ref{s:spectra}), 
 this extended diffuse emission is primarily thermal in nature whereas
 the cometary nebula, ring, and related features are dominated by 
 hard non-thermal spectra.

Analysis of these features is discussed 
 in greater detail in the following subsections
 beginning with an estimate of the proper motion of \cxou.
 
\subsubsection{Constraints on the Proper Motion of \cxoulong} \label{s:proper}

The current observations were taken 2582$\pm$1 days since the 37.5-ks observation
 ObsID 05531 was taken with \cxou\ at the aimpoint.
An earlier \cxo\ observation, ObsID 00760, was too short to be
 useful for precise measurements of proper motion.
Four point-like X-ray sources, in addition to \cxou, were identified common to both
 ObsID 05531 and the deeper of the two recent observations, ObsID 13736.
The spectra of three of these four sources have significant fractions of their
 counts above 2~keV indicating a high relative probability that they
 are distant background objects and hence fixed on the plane of the sky.
The fourth source is softer and possibly a foreground star. It is listed in
 the USNO-B1.0 Catalog  \citep{Mon03} with designation 1123-0129324 
 with a negligible proper motion.
Offsets (independently in R.A. and Decl.)
 between the two observations were computed by minimizing the error-weighted
 separations among the four sources. 
After applying these relative astrometric corrections, 
 the 99\%-confidence contour extremum sets an upper limit to the proper motion
 $\le$310$d_{1.5}$~km~s$^{-1}$ with the best-fit (but
 not statistically significant) direction toward the west.
Specifically, the estimated motion is 
 $-22.3\pm33.9$~mas~yr$^{-1}$ in R.A. and 
  $-0.1\pm33.9$~mas~yr$^{-1}$ in Decl.

\begin{center}
\includegraphics[width=0.9\columnwidth]{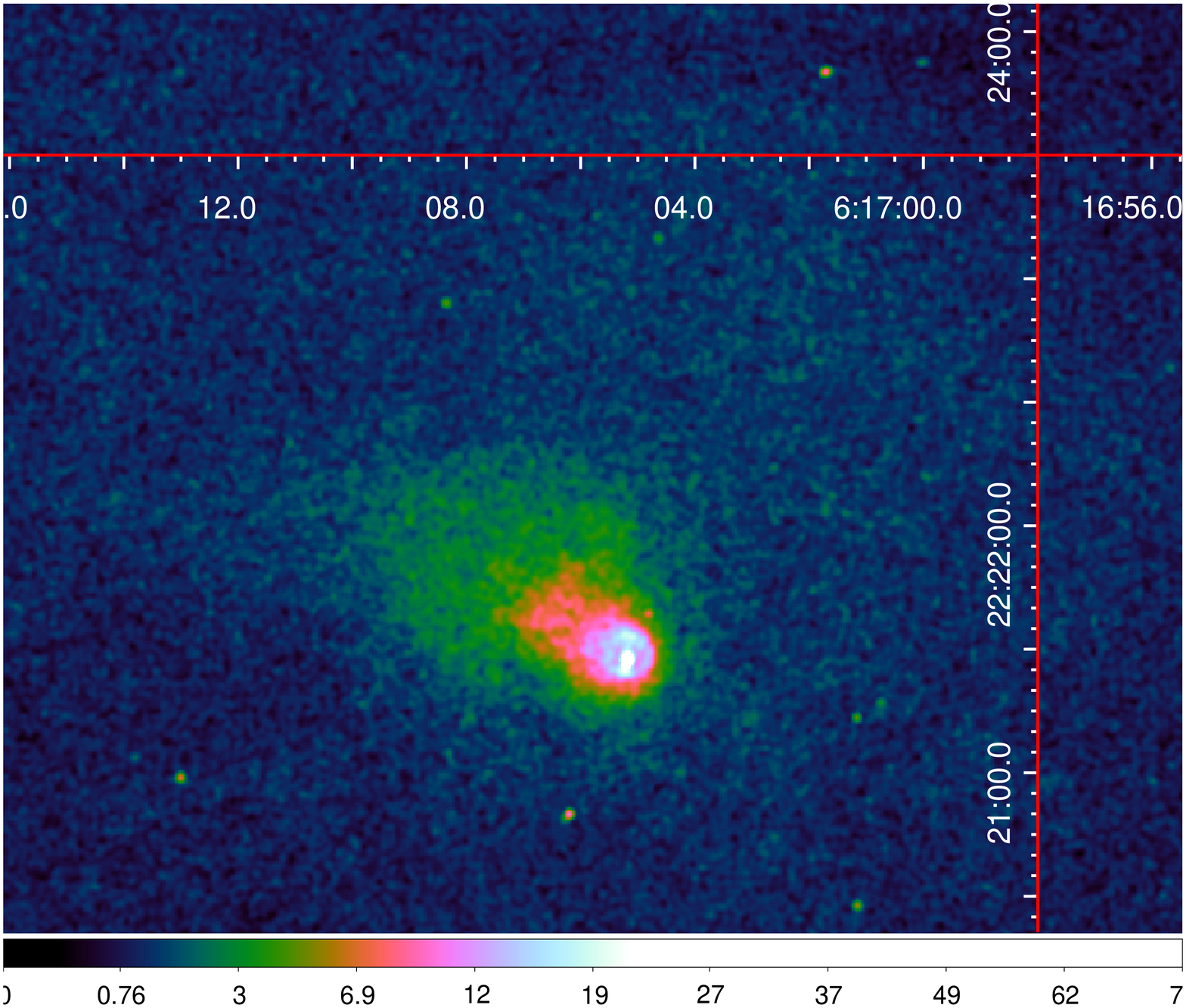}
\figcaption{X-ray image from the merged, 152~ks duration, \cxo-ACIS observation restricted to 
the 1.0$-$5.0~keV energy range and covering a 4.75\arcmin$\times$3.75\arcmin\ region near the
aimpoint and smoothed using a Gaussian kernel of 3 pixel (1.48\arcsec) radius. 
Colors are proportional to the number of X-ray
events per pixel using a square root scaling. \cxoulong\ is the bright 
white object surrounded by a ring (light blue) and the comet-shaped PWN (red and green).
Relatively low surface brightness structure (dark blue to black) is visible throughout
the S3 field of view.  
The center of the \ic\ supernova remnant lies $\sim$30\arcmin\ to the N of \cxou.
\label{f:fullS3image}}
\end{center}

\subsubsection{\cxou, the Jet, and the Ring: Subpixel Imaging} \label{s:ring}

Figure~\ref{f:fullS3image} displays the 2.0$-$4.0~keV subpixel-sampled region 
 around \cxou\ including the ring-like and jet-like structures.
This energy range best distinguishes the ring from other features
 and mutes the contribution from \cxou\ which is brightest at lower energies 
 (section~\ref{s:s_psr}).
Figure~\ref{f:subrp} shows the broad-band 
 surface brightness radial profile centered on the 
 position of \cxou, extending to 12.3\arcsec, and including only azimuthal
 angles between PA 200\arcdeg\ and 340\arcdeg\ ($\pm$70\arcdeg\ of West)
 in order to avoid contributions from the jet-like features.
This profile was obtained from the $1/2$-pixel sampled image. 
Also shown is a model point spread function for a monochromatic source
 at 2.5~keV generated using the Chandra Ray Tracer (ChaRT) and MARX simulation tools.
The ring appears as the peak at 4.8\arcsec\ with a (Gaussian) width of 1.2\arcsec.
These parameters were estimated by fitting the surface brightness profile with a combined power law plus Gaussian function.
Another peak in the radial profile appears between \cxou\
 and the ring. This feature is  about 1\arcsec\ from \cxou\ 
 but adding an additional Gaussian component shows this feature is not statistically significant.

\begin{center}
\includegraphics[angle=-90,width=0.9\columnwidth]{f03.eps}
\figcaption{X-ray brightness radial profile centered on \cxou\
 and restricted to azimuthal angles on the PA range 200\arcdeg\ to 340\arcdeg\
 ($\pm$70\arcdeg\ of due W).
The vertical scale is the number of counts per 
 0.246\arcsec$\times$0.246\arcsec\ ACIS CCD subpixel detected 
 in the 2.0$-$5.0~keV bandpass. The thin line is a ChaRT model
 of a 2.5~keV monochromatic point source. 
\label{f:subrp}}
\end{center}

The jet-like feature spans from the position of the pulsar northward and southward 
 until it appears to intersect the ring. 
The ring brightens at these intersections.
These intersections were described as compact emission components by 
 \citet{Gae06}.
The southern jet feature has a surface brightness about 75\% higher
 than the northern jet in the 0.5$-$8.0 keV energy range.

The center of the ring is not coincident with the location of \cxou.
A spatial model representing an annulus was fit to the sub-pixel X-ray image 
 in the 2$-$4~keV energy range after 
 ignoring data within a 2.5\arcsec\ radius of the center of the 
 ring (due to the brightness of the pulsar)
 and smoothing the data slightly (Figure~\ref{f:ring}). 
The model consists of two elliptical Gaussians plus a constant
 where one Gaussian has a positive amplitude and the other a negative amplitude 
 in order to approximate an elliptical annulus.
The location of the centroid of the ring model is 
 06$^{\rm h}$17$^{\rm m}$05.166$^{\rm s}$ 
 $+$22\arcdeg21\arcmin28.80\arcsec.
This location is offset by 2.7\arcsec\ (0.06\ddd\ ly) from \cxou\ at
 PA$\sim$340\arcdeg, i.e., the pulsar is roughly SSE of the ring center.
The ring is close to circular. 
The best-fitting parameters result in an aspect ratio of 0.86$\pm$0.05
 and an angular eccentricity, $\cos^{-1} (b/a) = 30.7 \pm 5.6$~deg
 where $a$ and $b$ are the major and minor Gaussian widths.
The PA is 84$\pm$3~deg; the ellipse major axis is nearly 
 in the east-west direction.

\begin{center}
\includegraphics[angle=-90,width=0.9\columnwidth]{f04.eps}
\figcaption{X-ray subpixel image of the ring surrounding \cxou\ restricted to
the 2.0$-$4.0~keV energy range and to the region within ten pixels ($\sim$2.5\arcsec) of the
center of the ring (including the pulsar) excluded. Solid contours indicate
counts pixel$^{-1}$ in the data and range from 1.05 to 1.55 in steps of 0.125.
Dotted contours indicate model fits to the ring morphology using the same
range of levels. The model is the sum of two elliptical Gaussians, one with a positive
amplitude and one with a negative amplitude (together approximating an annulus),
 and a constant. All parameters (centroids, widths,
amplitudes, and position angles of the Gaussians and the constant) were
allowed to vary in the fitting although the centroids of the two Gaussians were
 tied to the same values.
Circles are shown centered on the best-fit centroid (denoted by the asterisk)
and show that the overall shape of the ring is close to circular.
The position of \cxou\ is denoted by the red cross.
\label{f:ring}}
\end{center}

\subsubsection{Morphology of the Pulsar Wind Nebula} \label{s:bp}

Figure~\ref{f:bp} displays the 0.5$-$8.0~keV brightness profile
 along the major axis (PA 50\arcdeg) of the comet-shaped PWN. 
The profile has been averaged over a $\pm$30\arcsec\ region 
 perpendicular to this axis.
The peak of the profile coincides with the pulsar and offsets  
 upstream of the pulsar (toward the southwest) are taken as positive in this figure.
The ring is clearly visible as secondary peaks on either side of the pulsar. 
The profile $>$5\arcsec\ upstream of the pulsar 
 decreases smoothly out to at least 2\arcmin\ from \cxou\
 without any discernable structural features.
As with the upstream profile, the downstream profile first declines
 steeply from the exterior of the ring
 but then it reaches a broad peak extending from 15\arcsec\ to 25\arcsec\ away from the pulsar.
The profile then declines more slowly out to approximately 2\arcmin\ downstream
 without any additional features.

\begin{center}
\includegraphics[angle=-90,width=0.9\columnwidth]{f05.eps}
\figcaption{X-ray brightness profile along the major axis of the PWN.
Upstream is to the right.
The vertical scale is the number of counts per 
 0.492\arcsec$\times$0.492\arcsec\ ACIS CCD pixel detected 
 in the 0.5$-$8.0~keV bandpass in a 60\arcsec\ wide strip oriented 
 perpendicular to and centered on the major axis.
\label{f:bp}}
\end{center}

The X-ray image does show that the surface brightness distribution upstream of
 the pulsar is bow-shaped.
Parabolas were fit, by visual inspection,
 to the surface brightness contours in the upstream region of 
 a smoothed image (with the ordinate axis of the parabolas constrained to be  
 parallel to PA 50\arcdeg) as shown in Figure~\ref{f:parabolas}.
The parabolas are of the form 
 $(y/d_s)=0.54(x/d_s)^2+1$ where $d_s$
 is the distance from the location of \cxou\ (at the focus) 
 to the vertex of the parabola.

\begin{center}
\includegraphics[angle=-90,width=0.9\columnwidth]{f06.eps}
\figcaption{X-ray image of the upstream region around the PWN.
The image has been smoothed by three applications of a nearest-neighbor smooth.
Contours are drawn at levels of 1.5, 3.0, 5.0, and 10 c~pixel$^{-1}$.
Corresponding (color-matched) curves are parabolas 
 parameterized by visual inspection to represent the contours. 
The parabolas are constrained to have their ordinates parallel to the apparent
 line of motion of \cxou\ at PA 50\arcdeg. 
\label{f:parabolas}}
\end{center}

There is no evidence for a bow shock in either the \ha\ or \oiii\
 images of this region and
there is little or no correlation between the X-ray emission 
 in the upstream direction with the optical line emission. 
However, as a close-up image of the region shows (Figure~\ref{f:snrimage1}),
 there is \ha\ emission coincident with the jet-like X-ray feature.
This is likely an accidental coincidence as there are many similar \ha\
 features throughout the \ic\ SNR.
Similarly,
there is also a well-defined \ha\ wisp downstream of \cxou\
 that runs roughly along 
 the major axis of the PWN and thins in the downstream direction
 (Figure~\ref{f:snrimage1a}).
This feature is also visible in the [O\,III] image.
There is no X-ray feature correlated with this wisp.
Although its location is suggestive,
it is unclear if this \ha\ wisp is related to the PWN as there are 
 numerous wisps with similar morphology throughout the region that
 are obviously not related to the PWN proper.

\begin{center}
\includegraphics[angle=0,width=0.9\columnwidth]{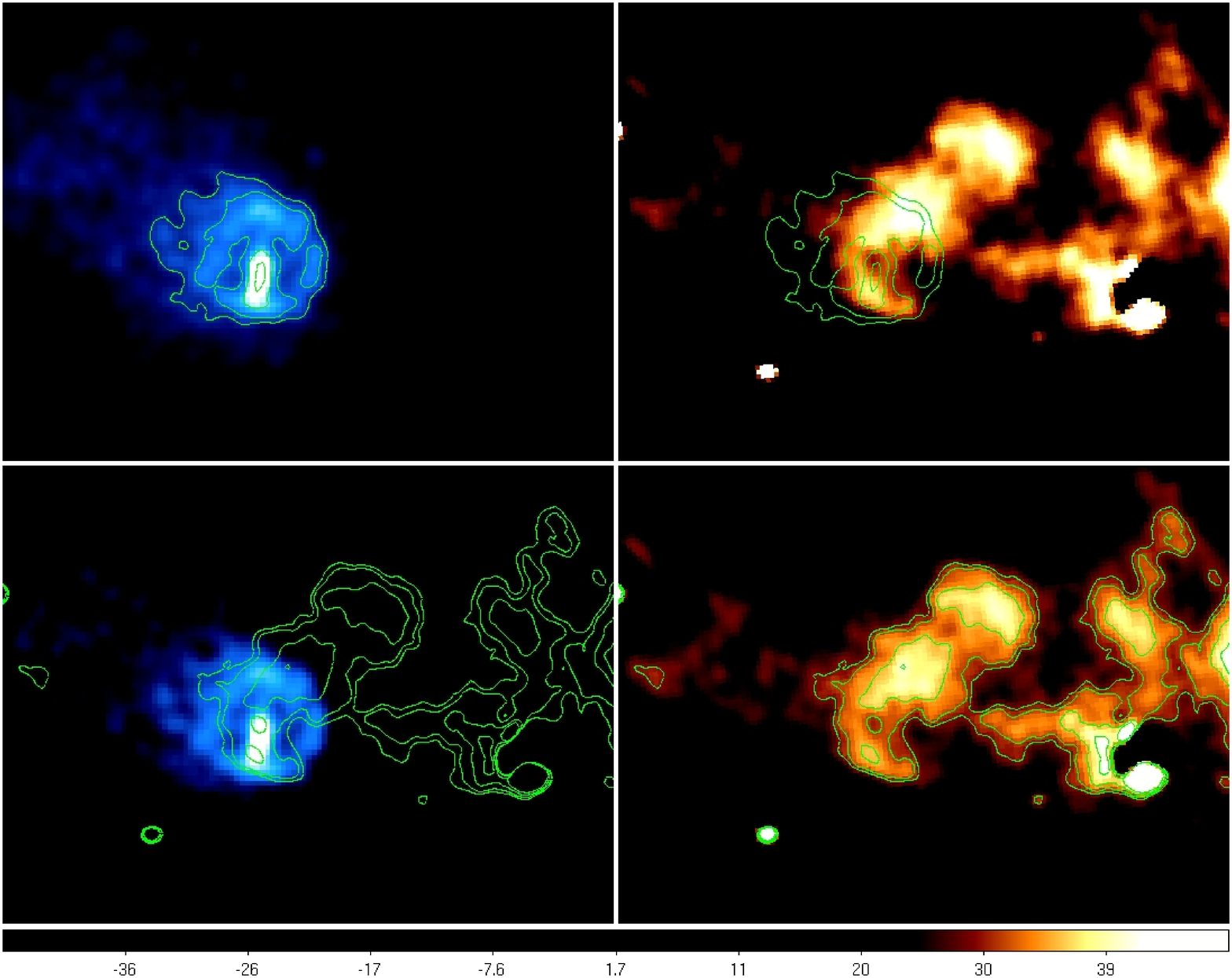}
\figcaption{
Detail of the ring and jet-like features in 
X-ray (left two panels) and \ha\ (right two panels)  
with X-ray contours (top panels) and \ha\ contours (bottom panels) overlaid.
Panels sizes are 1\arcmin$\times$0.75\arcmin. 
\label{f:snrimage1}}
\end{center}

\begin{center}
\includegraphics[angle=0,width=0.9\columnwidth]{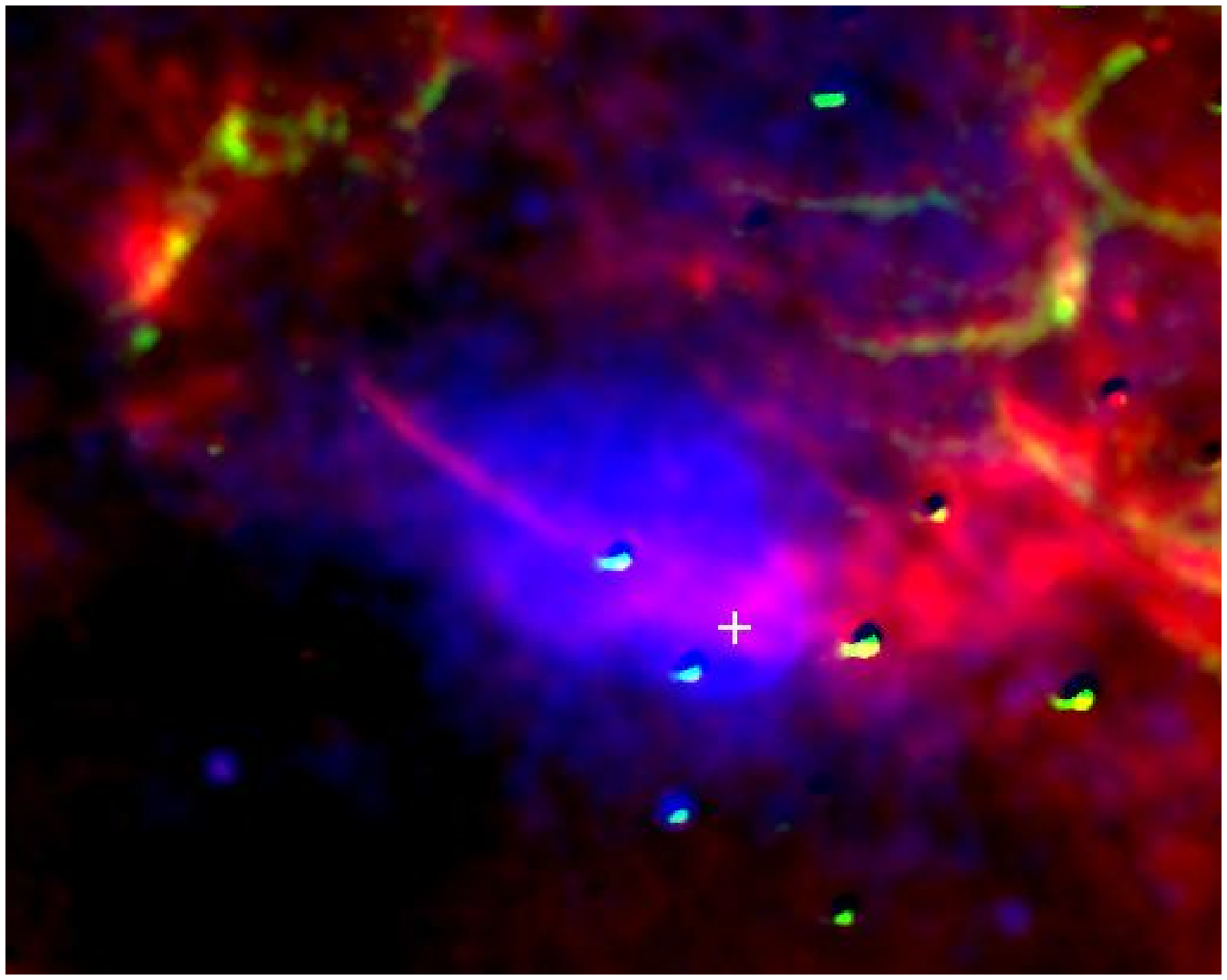}
\figcaption{
X-ray (blue), \ha\ (red), and \oiii\ (green) color 
composite of a 4.2\arcmin$\times$3.4\arcmin\ portion of the \ic\ field. 
The location of \cxou\ is marked with a white cross.
Note the \ha\ wisp extending east-west roughly along the line of apparent motion
 of \cxou\ superimposed on the PWN.
\label{f:snrimage1a}}
\end{center}

\subsection{X-ray Spectroscopy of Major Structures} \label{s:spectra}

\subsubsection{Regions Exterior to the PWN: Background SNR} \label{s:s_bg}

Customarily, a background region is selected from those regions remote from the
 primary source, in this case the PWN, in order to isolate the source spectrum
 for analysis. 
However, \cxou\ is located along the line of sight through the extended X-ray emission 
 associated with the SNR \ic\ and this emission is highly non-uniform as noted above.
Several regions were visually selected and analyzed to quantify the
 spectral properties of this ``background'' throughout the field of view of the S3 chip. 
These regions are numbered 1 through 9 in Figure~\ref{f:s3bg}
and their spectra are shown in Figure~\ref{f:bgspec}.
Assuming this emission is from hot plasma within the SNR,
the observed variations may be due to a combination of 
 different plasma temperatures, emission measures,
 elemental abundances, and intervening absorption.
The simplest model that can account for all these variables in XSPEC is an
 absorbed optically-thin thermal plasma model.
The spectral analysis reported in this section is limited to the 
 range 0.5$-$5.0~keV because of 
 the paucity of source photons above 5~keV.

\begin{center}
\includegraphics[angle=-90,width=0.9\columnwidth]{f10.eps}
\figcaption{Regions selected for spectral analysis of extended emission.
Numbering corresponds to the region numbers in Table~\ref{t:fitparams2}. 
Additional small rectangular regions upstream of \cxou\
are analyzed in section~\ref{s:s_upstream}.
The thick irregular region around \cxou\ delineates the outermost of 7 high-surface-brightness
 PWN regions investigated in section~\ref{s:s_pwn}
and the thin inner irregular region defines the radio-bright region 
 analyzed in section~\ref{s:r_pwn}. 
\label{f:s3bg}}
\end{center}

It was found that the spectra of these `outlier' regions
 cannot 
 be fit satisfactorily with a single-component absorbed (XSPEC model {\tt phabs}) 
 variable-abundance thermal emission ({\tt vmekal}) model.
A significant improvement could be made by adding either a second
 thermal component or a power law ({\tt powlaw}) component.
 
Resulting best-fit model parameters are listed in Table~\ref{t:fitparams2} for these 
 outlier fields. 
These include the absorption column density, $n_{\rm H}$,
 the power law index, $\Gamma$,
 the thermal component temperature, $kT$,
 the model component normalizations, $K$\footnote{
$K_{\rm POW}$ is in units of flux density, ph~keV$^{-1}$~cm$^{-2}$~s$^{-1}$ at 1 keV and
$K_{\rm THM}$ is in units of $10^{-14} (4 \pi d^2 )^{-1}$ times the emission measure, in cm$^{-3}$. },
 the 0.5$-$5.0~keV flux, $f$, in \ergcms\ (uncorrected for absorption)
 and the $\chi^2$ fit statistic and number of degrees of freedom, dof.
Here and elsewhere errors are 90\% confidence extremes for a single interesting parameter
 unless otherwise noted.

\begin{center}
\includegraphics[angle=-90,width=0.9\columnwidth]{f11.eps}
\figcaption{Observed spectra of the 9 numbered regions shown Figure~\ref{f:s3bg}.
The spectra have been binned up to 10 channels in order 
 to achieve a minimum detection significance of 10 for display purposes.
Colors correspond to regions 1:cyan, 2:gray, 3:magenta, 4:blue, 5:black,
 6:green, 7:red, 8:orange 9:dark gray as defined in Figure~\ref{f:s3bg} and Table~\ref{t:fitparams2}.
Note the high flux above $\sim$2.0~keV in the spectrum 
 of regions 7$-$9 indicating a strong hard spectral component.
The different relative strengths of emission lines among the various
 regions are due, in part, to real differences in elemental abundances
 as the best-fitting models all have similar 
 plasma temperatures and absorption columns (Table~\ref{t:fitparams2}).
\label{f:bgspec}}
\end{center}

The first thermal component temperature is typically $kT\sim0.5$ to 0.8~keV.
It was found that the second component 
usually had a steep power law slope 
 (photon index $\Gamma \gtrsim 2.5$) or only a slightly higher temperature 
 ($kT\sim0.8$ to 1.0~keV) suggesting  that this component may
 represent the 
 range of thermal emission temperatures expected along a given line of sight through the SNR
 rather than a true non-thermal component typically indicated by a power law shape.
This is true of most of these outlier regions except regions 7$-$9 where  
 $\Gamma < 2.3$ (or $kT_2>3.5$~keV). In these cases, it is likely that
 there is truly a hard non-thermal component even if the two-$kT$ model
 is formally a better fit as such high temperatures are atypical of SNR spectra.
The PWN likely contributes this non-thermal emission component
 in regions 8 and 9 because they are immediately adjacent to the high-surface-brightness PWN.

The elemental abundances were not well constrained in any of these models 
 but strong emission lines of 
 H-like O and Ne, and He-like blends of Ne, Mg, Si, S are present throughout
 as shown in Figure~\ref{f:bgspec} whereas Fe emission is weak.
This is indicative of a Type~II core collapse SN origin.

Overall,
since the spatial scale of the observed brightness variations are of order
 1\arcmin, and the spectra of the different regions sampled also vary,
 none of these regions (nor any other region on S3) can be safely
 defined as a background to be used in fitting spectra of the PWN.
Thus, the ``blank sky'' 
 background\footnote{obtained from the CalDB and accessed following the 
 CIAO science 
 thread http://asc.harvard.edu/ciao/threads/acisbackground/}
 appropriate to this observation is the 
 only background used in subsequent spectral analysis
 with the exception of the pulsar, \cxou, and the jet-like and ring features.
For these objects, the space between the pulsar and ring is relatively faint and
 local backgrounds were chosen from this region. 

\subsubsection{Spectra and Timing of \cxoulong} \label{s:s_psr}

The spectrum of \cxou\ was extracted from a 2-pixel (0.984\arcsec) radius circle centered at
 the location of the pulsar (see also Figure~\ref{f:subrp}).
A nearby background was obtained by combining data from five 2-pixel radius circles
 located interior to the ring and excluding the jet-like features.
This results in an estimated 925$\pm$35 net source counts in the 0.5$-$8.0~keV energy range. 
Absorbed blackbody and neutron star atmosphere 
 \citep{Pav95,Zav09} models 
 (plus a power law component)
 were fit to the spectrum over the 0.5$-$8.0~keV energy range
 after grouping to ensure a minimum of 20 background-subtracted events per 
 spectral energy bin.
The spectrum, models, and fit residuals are shown in Figure~\ref{f:psrspec}.
The neutron star atmosphere model assumed a distance to the star of 1.5~kpc, a 
 radius $R_{\rm NS}=10$~km, a mass $M_{\rm NS}=1.4$~\msun\
 (corresponding to a gravitational redshift parameter 
 $g_r \equiv [1 - 2 G M_{\rm NS}/(R_{\rm NS} c^2)]^{1/2} = 0.766$)
 and a magnetic field strength of 
 10$^{13}$~G.

\begin{center}
\includegraphics[angle=-90,width=0.9\columnwidth]{f12a.eps}
\includegraphics[angle=-90,width=0.9\columnwidth]{f12b.eps}
\figcaption{Observed spectrum of \cxoulong\ extracted from within an $\sim$1\arcsec\ 
 (2-pixel) radius circle. A background, extracted from small regions surrounding the 
 source, interior to the ring, and excluding bright regions associated with the jet-like
 feature, has been subtracted.
The spectrum has been grouped to obtain a minimum of 20 counts per spectral bin.
The solid line denotes the full model 
 ({\sl top:} absorbed blackbody plus power law;
  {\sl bottom:} absorbed neutron star atmosphere plus power law),
The dotted lines are the contributions of the power law component and the dashed lines
 are the blackbody ({\sl top}) and NSA model ({\sl bottom}) components.
The lower panels show the respective fit residuals.
\label{f:psrspec}}
\end{center}

The best-fitting model parameters, 
 derived properties, and fit statistics are  listed in Table~\ref{t:fitparams1},
 where $T^{\infty}$ and $R^{\infty}$ are the effective temperature and radius as measured by a 
 distant observer, and $L_{\Gamma}$ is the 0.5$-$8.0~keV luminosity of the non-thermal (power law)
 component. Errors in Table~\ref{t:fitparams1} are at the 1$\sigma$ (68\%) confidence level.
Both models provided acceptable fits to the data.
These spectral properties are typical for a cooling neutron star
 although the blackbody model radius, $R^{\infty} = 1.63$~km, implies emission from a small hotspot
 rather than the entire neutron star surface.
They indicate a bolometric luminosity of (1$-$3)$\times$10$^{32}$~\ergl.

A search for a possible periodicity in the X-ray flux from \cxou\ was
 restricted to periods $P>6.5$~s due to the 3.24-s time resolution of the ACIS instrumental setup.
The Rayleigh test yielded a  maximum power at a period of 31.18~s.
Applying the appropriate probability distribution
\citep{Gro75} sets an upper limit to the power at 99\% confidence 
 that translates to an upper limit on the pulsed fraction of 36\% assuming a sine-wave signal and accounting for the background.
 
\subsubsection{Spectra of the Jet and Ring} \label{s:s_ring}

The spectrum of the southern jet-like feature was extracted from a 
 10\arcsec$\times$5\arcsec\ region oriented N$-$S and centered at 
 06$^{\rm h}$17$^{\rm m}$5.25$^{\rm s}$ 
 $+$22\arcdeg21\arcmin25.5\arcsec.      
The background-subtracted source spectrum (using the same background regions as for \cxou) 
 was grouped to a minimum of 10 events per spectral energy bin and fit using 
 an absorbed power law. 
There were too few counts (233$\pm$21) to justify attempting more complex models.
The best-fitting model parameters are $n_{\rm H}=6.3^{+3.0}_{-1.2} \times 10^{21}$~cm$^{-2}$ and
 $\Gamma=1.2^{+0.4}_{-0.3}$ ($\chi^2/{\rm dof}=20.9/18$) resulting in an 
 absorbed 0.5$-$8.0~keV model flux of 2.1$^{+2.8}_{-1.3}\times 10^{-14}$~\ergcms\ and
 absorption-corrected luminosity of 7.0$^{+8.0}_{-4.0}\times 10^{29} d_{1.5}^2$~\ergl.
There were too few counts for spectral analysis of the northern portion of the jet.

The spectrum of the ring was extracted from a circular annulus 
 lying between radii of 3.7\arcsec\ and 5.9\arcsec\ from the ring center.
An absorbed power law model with parameters 
 $n_{\rm H}=5.2^{+0.6}_{-0.5} \times 10^{21}$~cm$^{-2}$,
 $\Gamma$=1.4$\pm$0.1 produced an acceptable fit to the spectrum
 ($\chi^2/{\rm dof}=194.3/182$).
The resulting 
 0.5$-$8.0~keV model flux is (3.1$\pm$0.3)$\times$10$^{-13}$ \ergcms\
 and the absorption-corrected luminosity is 1.1$\times$10$^{32} d_{1.5}^2$~\ergl.

\subsubsection{Spectra Upstream of the Pulsar Wind Nebula} \label{s:s_upstream}

Spectra were extracted from a series of regions located
 upstream of the pulsar to search for a spectral transition between the non-thermal
 shocked PWN emission and thermal emission from shocked ambient medium further
 upstream. 
Figure~\ref{f:s3bg} shows the locations of these regions.
They are 40$\times$100 pixel$^{2}$ (0.269~arcmin$^2$) rectangles oriented with the long side
 perpendicular to the line of apparent motion of the pulsar.
Table~\ref{t:fitparams3} presents the spectral fitting results for these regions
 (ordered left to right by increasing distance from \cxou).
As with the outlier regions discussed in section~\ref{s:s_bg}, absorbed 
 two-component models were applied. 
Both models provide similarly acceptable fits although the thermal plus power law 
 is formally the superior model.
With the exception of the region nearest to \cxou,
 the temperature of the soft thermal component in all cases,
 $kT\sim0.6-0.7$~keV, is consistent with the 
 soft component deduced for the outlier regions of section~\ref{s:s_bg}.
Most regions have an obvious hard component ($\Gamma \lesssim 2.3$ or $kT_2>3.5$~keV)
 that is poorly modeled by a second {\sl thermal} component.
For the non-thermal power law (plus soft thermal) models, the trend is for the power law index 
 to steepen and both the power law flux and its fraction of the total flux to decrease
  with distance from \cxou. 
This is certainly true of the innermost three regions
 although the trend is less evident in the outer three regions where the 
 power law component is much weaker though still required in the modeling. 
While there is no abrupt spectral transition from non-thermal- to thermal-dominated
 emission, a non-thermal component is clearly present at least out to 2\arcmin\ upstream from
 \cxou, which is consistent with the smoothly-varying 
 surface brightness distribution of this upstream region 
 (cf. Figure~\ref{f:bp} and section~\ref{s:bp}).

\subsubsection{Spectra of the Pulsar Wind Nebula: Contour Binning} \label{s:s_pwn}

The contour binning method of \citet{San06} was applied to a 
 region containing the pulsar, PWN, and immediate surroundings
 as described in \S\ref{s:dataprep}. 
By definition, this binning method traces the decrease in surface brightness with distance 
 from \cxou.
Using this prescription, the PWN is subdivided into roughly
 concentric ``annuli''  with the innermost region 
 containing \cxou, the ring and the jet.
The outermost of these regions is indicated in Figure~\ref{f:s3bg} by the thick irregular 
 contour surrounding \cxou.
By design, each region contains roughly 10$^4$ X-ray events.
Because the extended thermal emission from the \ic\ SNR is so spatially non-uniform,
 the contour binning method did not satisfactorily bin the regions beyond the PWN. 

The results of model fits to the spectra of these regions are given in 
 Table~\ref{t:fitparams4} (a 3-pixel (1.48\arcsec) radius circle centered on \cxou\ was excluded from the 
 innermost region, denoted Region 1 for spectral analysis; regions 2 through 7 lie sequentially ourward).
In addition to model parameters and derived properties 
 listed previously (Table~\ref{t:fitparams2} and ~\ref{t:fitparams3}), Table~\ref{t:fitparams4}  includes fluxes, $f^{\rm cor}$,
 corrected for absorption along the line of sight.
The spectra were fit on the extended 0.5$-$8.0~keV energy range and
 fluxes listed in Table~\ref{t:fitparams4} are over this same interval.
 
An absorbed power law model provided a statistically-superior fit to the spectra of the
 three innermost regions.
An absorbed two-component ({\tt powerlaw} and {\tt vmekal} or two {\tt vmekal}) model
 was needed for the outer zones but one component is always hard
 ($\Gamma < 2$ or $kT>5$~keV) and dominates the 0.5$-$8.0 keV flux;
 again, indicating a true non-thermal component is present as argued 
 in section~\ref{s:s_bg}.
For this reason, the two {\tt vmekal} component model is not considered further.
The soft thermal component in the two-component non-thermal model is 
 generally cooler, $kT\lesssim 0.4$~keV, throughout these PWN regions than in the `outlier' fields
 (Section~\ref{s:s_bg}) but note that the soft component   
 contributes at most 5\% of the flux in all but the 
 outermost PWN region.
The outermost region is similar to the outlier regions 7$-$9 (Table~\ref{t:fitparams2}) in that its 
 soft thermal component temperature is 0.67~keV and its power-law index is $\sim$2.
Unlike these outlier regions, nearly all the flux in even this outermost PWN region is provided by the 
 non-thermal model component.

To test the significance of the low temperature in the inner zones, the fits were repeated
 with the {\tt vmekal} model temperature parameter fixed to its value in the outermost zone,
 $kT=0.67$~keV.
This results in some additional flux in the 1-2 keV band 
 which forces the power law index to flatten slightly to compensate. 
However, this is only a modest change in slope and within the errors 
 obtained if the thermal model temperature were left as a free parameter.
For example, in the fourth zone, $\Gamma$ decreases from 1.9$\pm$0.1 to
 1.8$\pm$0.1.

\begin{figure*}
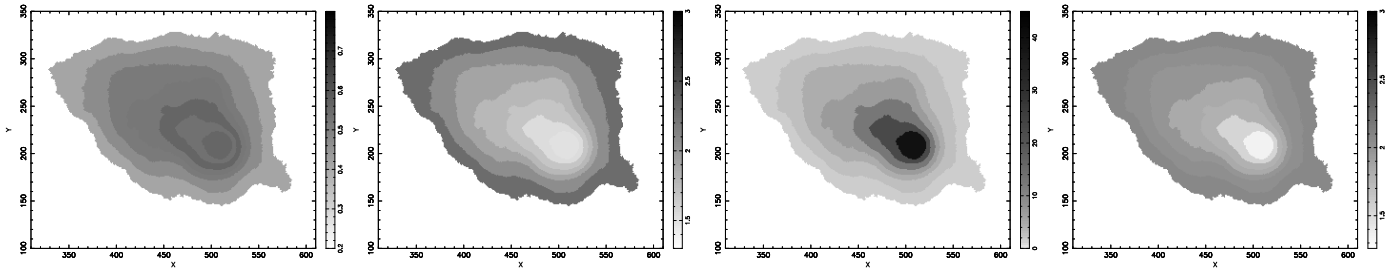

\begin{center}
\includegraphics[angle=-90,width=4.5cm]{f13a.eps}
\includegraphics[angle=-90,width=4.5cm]{f13b.eps}
\includegraphics[angle=-90,width=4.5cm]{f13c.eps}
\includegraphics[angle=-90,width=4.5cm]{f13d.eps}
\figcaption{
Spectral properties of the seven regions encompassing the PWN as defined
 by the contour binning method of \citet{San06} and listed in Table~\ref{t:fitparams4}.
Properties are shown for the single absorbed power law model.
Region numbers listed in Table~\ref{t:fitparams4} column headings (1$-$7)
 are ordered from the inside outward.
Each region encompasses an area of approximately uniform surface brightness.
The innermost (region~1) is approximately 15 pixels ($\sim$7.5\arcsec) in radius and
  contains the ring, and the jet features (\cxou\ is excluded); 
 region~2 encompasses the ``plateau'' region (see Figure~\ref{f:bp}).
The panels display ({\sl clockwise from top left:}) 
 the fitted absorption column density, \nh/10$^{22}$~cm$^{2}$, 
 the power law model photon index, $\Gamma$, 
 the observed 0.5$-$8.0~keV intensity, $\epsilon_{\rm X}/10^{-16}$~\ergcms~arcsec$^{-2}$,
 and the power law model normalization, 
 $K_{\rm POW}/10^{-4}$~photons~kev$^{-1}$~cm$^{-2}$~s$^{-1}$ at 1~keV.
\label{f:contbinmap}}
\end{center}
\end{figure*}

Maps of several of the resulting best-fitting model parameters 
 are displayed in Figure~\ref{f:contbinmap} for the single-component absorbed {\tt powerlaw} model.
These maps show that the power law slope smoothly increases with distance 
 from \cxou, that the intensity decreases, and that the best-fitting 
 column density is roughly constant but decreases in the outermost zone.
This latter trend is perhaps influenced by the simplicity of this single-component model.

\subsubsection{Spectra of the Pulsar Wind Nebula: Comparisons to Radio Data} \label{s:r_pwn}

The X-ray emission in the \cxou\ pulsar wind nebula is clearly dominated by a non-thermal component 
 suggesting a synchrotron emission source.
Radio images of the region \citep{Olb01, Cas11} also
 show a cometary morphology 
 which may imply a similar origin.
However, the radio and X-ray data have much different spatial resolution and sensitivity
 so that a direct comparison is non-trivial.
Specifically, the signal to noise of the higher frequency radio data 
 is too low to allow for measurements in small regions for comparison with the X-ray fluxes.
Instead, by cutting the upper and lower uv range of the 4.8~GHz and 8.4~GHz measurements
 to a common $\sim$1.7\arcmin$\times$0.77\arcmin\ region 
(see Figure~\ref{f:s3bg}) 
 with equal coverage at both frequencies (and at 330~MHz),
 radio spectral index measurements of roughly the whole \cxou\ PWN can be obtained.
Comparison of the coverage after cutting showed no apparent gaps in the coverage 
 and thus both frequencies should be sensitive to emission over a similar range of spatial scales. 
The data were therefore re-imaged and convolved with a circular beam of size
 8.5$^{\prime\prime}$ and the images were blanked for emission falling below a 
 three sigma cutoff before measurements were taken of the total flux of this region. 
This yields nearly equal fluxes at all three frequencies,
 84.3$\pm$7.6, 87.6$\pm$3.0, and 87.0$\pm$0.6~mJy at
 330~MHz, 4.8~GHz and 8.4~GHz, respectively, and therefore a flat spectral
 index of $\alpha_{\rm R}=0.01$ in the radio band
 consistent with \citet{Olb01} and \citet{Cas11}. 
 
The combined 330~MHz image does not suffer from these sensitivity issues but does not have  
 the high spatial resolution of the X-ray data.
Therefore, instead of the contour binning regions of section~\ref{s:s_pwn},
 a set of successively larger elliptical annuli 
 approximating the innermost five contour binning regions
 was analyzed at both 330~MHz and X-rays.
As before, the X-ray spectrum is predominantly non-thermal
 and a single absorbed {\tt powerlaw} model provided an acceptable fit.
Here, \cxou\ was {\sl not} excluded since it cannot be cleanly masked out of the low-resolution radio image.
The 330~MHz radio fluxes were calculated by integrating over each region after 
 subtracting an estimated contribution due to emission from the surrounding SNR
 as noted in section~\ref{s:dataradio}.
Table~\ref{t:Ellipsesfitpars} summarizes the geometry 
 and the radio and X-ray properties for these regions.

If the SNR-background-subtracted 330~MHz radio flux
 and the non-thermal (power-law component) X-ray emission
 arise from the same physical mechanism,
then their spectral shapes should be simply related.
The slopes of these spectra are obviously different: 
 The radio spectrum, $\alpha_{\rm R}$, is flat, at least on average over the entire region, 
 while $\alpha_{\rm X} \equiv \Gamma-1 \simeq 0.5-1$ depending on the (elliptical) region.
For a power-law X-ray spectrum with a photon index $\Gamma$ 
 and normalization $K_{\rm POW} = 10^{-4} K_{-4}$ photons (cm$^2$~s~keV)$^{-1}$ at $E=1$~keV,
 the energy flux spectrum is
 $F_{\nu,\rm X} \propto K_{-4} \nu^{-\Gamma+1}$.
Extrapolation of this X-ray spectrum towards lower frequencies 
 intersects the extrapolation of the radio spectrum, 
 $F_{\nu,\rm R}=F_{\rm 1 GHz}(\nu/{\rm 1 GHz})^{-\alpha_{\rm R}}$,
 at frequency $\tilde{\nu}$ when
\[
(\alpha_{\rm X}-\alpha_{\rm R}) \log \tilde{\nu} = \log ( K_{-4} / F_{\rm 1 GHz} ) + 8.3838\alpha_{\rm X} - 4.1798,
\]
where $\tilde{\nu}$ is in units of GHz and $F_{\rm 1 GHz}$ is in mJy.
Inserting the appropriate values from Table~\ref{t:Ellipsesfitpars},
 and assuming $\alpha_{\rm R}=0$ everywhere (and no uncertainty in the radio flux densities), 
 the estimated frequencies  $\tilde{\nu}$ are as listed in the last row of Table~\ref{t:Ellipsesfitpars}.
In the X-ray hard (flat) inner elliptical regions 1 and 2,
 the extrapolation crosses the radio spectrum at frequencies lower than 4.8 or 8.4~GHz
 which means that 
there should be at least two breaks in a power law spectrum  
 or the radio spectrum is not flat in those regions.

\section{Discussion} \label{s:discussion}

\subsection{Properties of the Central Source} \label{s:centralsource}

Deep \cxo/ACIS observations show the point-like object, \cxoulong,
 has an X-ray spectrum 
dominated by a thermal component with parameters  
 typical of 
 a cooling neutron star. The fit with a hydrogen 
atmosphere (NSA) model
 shows an effective temperature 
$T_{\rm eff}^\infty \approx 6.8\times 10^5$ K
and a bolometric
luminosity $L_{\rm bol}^\infty \approx 2.6\times 10^{32}$ erg s$^{-1}$
(see Table \ref{t:fitparams1}). 
Such temperature and luminosity 
are very close to those obtained from the NSA fit of the thermal
component of the Vela pulsar 
\citep{Pav01}, whose age $\tau$ is about 20--30 kyr, larger than
the characteristic spin-down age $\tau_{\rm sd}\equiv P/2\dot{P} =11$ kyr 
\citep{Lyn96}. Similar temperature and luminosity for $\tau\sim 30$ kyr
were obtained in some neutron star cooling models \citep[e.g., ][]{Pag09}.
Therefore, we can expect that age of the neutron star in \cxou\ is a few times
$10^4$ yr. It is consistent with the IC\,443 age $\sim 30$ kyr estimated by
\citet{Che99}, although some authors inferred much younger ages for
this SNR (e.g.,
\citet{Pet88} suggested $\tau\sim 3$ kyr). 

Compared to the NSA model, the blackbody fit of the thermal component gives a higher temperature,
$T_{\rm BB}\approx 1.5\times 10^6$ K, and a smaller radius, $R_{\rm BB}\approx 1.6 d_{1.5}$ km,
similar to the Vela pulsar.
  Such emission could emerge from a hot spot
on the neutron star surface heated by precipitation of relativistic
particles from the pulsar magnetosphere,
 or it could reflect strong nonuniformity of
the surface temperature due to anisotropy of heat
conductivity of the neutron star crust caused by very strong magnetic
fields \citep[see, e.g., ][]{GepVig14}. The non-detection of 
pulsations with $P>6.5$ s does not contradict the small size of the
emitting region because such young pulsars have much shorter 
periods\footnote{Periods 
 larger than 6.5 s are observed in magnetars, but this neutron
star does not show any magnetar properties.}. 

Since the NSA and blackbody models yield fits of about the same quality, we cannot
prefer one model to the other based solely on the observations. However, 
the NSA fit looks somewhat more preferable because it gives reasonable
cooling neutron star properties at the expected SNR age and distance.

The detection of the nonthermal power-law component
in the neutron star spectrum and the presence of the non-thermal X-ray
nebula around the neutron star show that \cxoulong\
is a rotation-powered pulsar.
The nonthermal component, presumably emitted from the pulsar's
magnetosphere, dominates the pulsar spectrum at $E\gtrsim 1.7$ keV.
The spectral slope 
and the ``isotropic luminosity'' of the nonthermal component
($L_\Gamma \equiv 4\pi d^2 f_{\rm pl}^{\rm cor}$)
are also close to those of the Vela pulsar\footnote{Note a typo
in \citet{Pav01}: The power-law luminosities in Table 1 should be in
units of $10^{31}$ (not $10^{32}$) erg s$^{-1}$.}, for both thermal 
component models used in fit. 

Since pulsations have not been detected, we do not know the
pulsar's spin-down properties, $\dot{E}$ and $\tau_{\rm sd}$. We, however,
can 
constrain them from a comparison of the PWN 0.5--8 keV 
luminosity
$L_{\rm pwn}\approx 
1.4\times 10^{33} d_{1.5}^2$ erg s$^{-1}$
 (given by the sum of the non-thermal components of the 7 contour-binning regions
 of Section~\ref{s:s_pwn}) 
with the X-ray luminosities of PWNe produced by
 pulsars with known $\dot{E}$ and $\tau_{\rm sd}$.
Assuming $d_{1.5}=1$ and using Figure 5 from \citet{KarPav08},
Figure 2 from \citet{Li08}, and Figure 3 from \citet{Kar12}, 
we obtain a range of possible spin-down luminosities 
$\dot{E}\approx (1$--$30)\times 10^{36}$ erg s$^{-1}$, with the most likely
value around $3\times 10^{36}$ erg s$^{-1}$. Similarly, using Figure 5 from
\citet{Li08}, we obtain a range of possible spin-down ages, $\tau_{\rm sd}
\sim 10$--100 kyr, with the most likely value around 20 kyr, consistent with
the age estimates from the properties of the thermal component. 
From these estimates
we obtain constraints
on the period,
$P\sim 0.1$--0.6 s,
and the
surface magnetic field, $B\sim (0.4$--$4)\times 10^{13}$ G, with most likely
values around 0.25 s and $2\times 10^{13}$ G, respectively.

Note that the true pulsar's X-ray luminosity is apparently much
higher than the above-estimated $L_\Gamma$ because the (distance-independent)
 ratio 
$L_{\rm pwn}/L_\Gamma\sim 100$--200 is much larger than for any known pulsar-PWN
pair \citep[see Figure 5 in ][]{KarPav08}. This suggests that the pulsar
beam misses the Earth, which could also be the explanation for
 non-detection of the pulsar
in the radio and $\gamma$-rays.

\cxou\ is separated from the apparent center of the \ic\ SNR by 15\arcmin\
 in a direction nearly due south.
This corresponds to a projected length $s_{\perp} = 2\times10^{19} d_{1.5}$~cm 
 and a projected velocity $v_{\perp} = 
640\,d_{1.5} \tau_{4}^{-1}$~\kms.
This is below the observed upper limit (Section~\ref{s:proper}) 
 of $310\, d_{1.5}$~\kms\ if $\tau_4\equiv \tau/10^4\,{\rm yr}\gtrsim 2$,
 which is consistent with the possible IC\,443 age and the
spin-down age estimated above 
 from the empirical 
$L_{\rm pwn}$--$\tau_{\rm sd}$ correlation.
The true speed is $v_{\rm psr} = v_{\perp} (\sin i )^{-1}$,
 where $i$ is the angle between 
 the direction of motion and the line of sight.
Observed pulsar speeds range up to $\sim$1000~\kms\ 
\citep[e.g., ][]{Arz02}
 which constrains the angle: $\sin i \gtrsim 0.64 \tau_4^{-1} d_{1.5}$,
assuming \cxou\ was born in the center of \ic\ (e.g., $i\gtrsim 19^\circ$ for
  $\tau_4 \sim 2$, $d_{1.5}=1$).
The angle $i$ is larger for more typical pulsar velocities -- for instance,
$32^\circ \lesssim i\lesssim 53^\circ$ for $v=400$ km s$^{-1}$ and a reasonable
age range $2\lesssim \tau_4 \lesssim 3$, at $d_{1.5}=1$.

\subsection{The surrounding medium} \label{s:surround}

All regions imaged by \cxo\ surrounding \cxou\ and its PWN
 have a strong thermal emission component,
likely emitted from the \ic\ SNR.
This 
likely means that \cxou\ lies within this warm plasma, 
 but a simple line-of-sight projection cannot be ruled out
 (and, thus, the distance to \cxou\ remains unknown).
The typical temperature of this warm plasma is $kT \sim 0.7$~keV (Section~\ref{s:s_bg})
 corresponding to a sound speed,
 $c_s = (\gamma kT/\mu m_{H})^{1/2} \approx
430\,(kT/0.7\, {\rm keV})^{1/2} (0.6/\mu)^{1/2}$ km s$^{-1}$
 for a $\gamma = 5/3$ gas and molecular weight $\mu\simeq 0.6$, appropriate for 
 fully-ionized medium with roughly cosmic abundances.
Thus, if \cxou\ is within this plasma and not just lying along the same line of sight,
 then the pulsar motion cannot be highly supersonic; the Mach number
 ${\cal M} = v_{0}/c_s \lesssim 2.3$ for $v_{0} < 10^3$~\kms.
Here, $v_{0}$ is the speed with respect to the ambient medium;
 $v_{0} \sim v_{\rm psr} $ in the absence of any bulk motions in the plasma.
Whether the speed is supersonic, transonic, or subsonic has a strong bearing on 
 the shape of the nebula (which is also influenced by its orientation relative to 
 the line of sight) and the properties of the particle
 and electromagnetic field flow within the nebula.
This is addressed in more detail in Section~\ref{s:mach}.

Assuming, for the present, that \cxou\ and its PWN are embedded in the \ic\ SNR,
 the ambient thermal pressure, $p_{\rm thm} = \rho (kT/\mu m_H)$,
 where $\rho$ is the ambient mass density,
 can be estimated with the aid of the {\tt vmekal} model fit parameters.
The {\tt vmekal} model norm is 
$ 
K_{\rm THM} = 10^{-14} (4 \pi d^2 )^{-1} \int n_e n_H dV 
       \simeq 10^{-14} (4 \pi d^2 )^{-1} n_H^2 V
$, 
where $n_H$ is the hydrogen number density,
 and 
$V$ is the volume of the emitting region. The volume can be expressed in terms of
the area $A$ of the region image (in arcmin$^2$) and the average depth $l_{19}$
(in units of $10^{19}$ cm) of the region along the line of sight:
$V=1.8\times 10^{55} A l_{19} d_{1.5}^2$ cm$^3$.
Using this expression for $V$ and assuming that fully ionized hydrogen gives the main contribution,
the hydrogen number density and the 
 ambient thermal pressure can thus be estimated as follows:
\begin{eqnarray*}
n_H &=& 
0.39\, K_{-4}^{1/2}(A l_{19})^{-1/2}\,\,\,{\rm cm}^{-3}\,,  \\ 
p_{\rm thm} & \simeq & 2 n_H kT = 0.92 \times 10^{-9} K_{-4}^{1/2}(A l_{19})^{-1/2}\,kT\,\,\, {\rm dyn}\, {\rm cm}^{-2}\,,
\end{eqnarray*}
where $K_{\rm THM}=10^{-4} K_{-4}$ and $kT$ is in units of keV.

The $n_H$ and $p_{\rm thm}$ estimates are proportional to $l_{19}^{-1/2}$,
which depend on the SNR geometry and the J0617 position within the SNR.
\cxou\ lies, in projection, about $15'$ ($2\times 10^{19} d_{1.5}$ cm)
from the center
of the 
$\sim 20$\arcmin\ radius \ic\ remnant
 (assuming a spherical remnant).
Thus, a chord through the remnant along the line of sight to \cxou\ 
should have a length of about $2\times 10^{19}$ cm 
or less, so typical expected values are
$l_{19} \sim 1$ 
(corresponding to $\sim 7.5\arcmin$ in the sky plane).

Using the values from spectral fits to the nine outlier regions, Table~\ref{t:fitparams2},
 the mean hydrogen densities (and standard deviations)
 for the low- and high-temperature {\tt vmekal} components 
 are $n_H / l_{19}^{-1/2} = 0.5 \pm 0.1$ 
 and $0.6 \pm 0.2$~cm$^{-3}$.
The corresponding pressures are 
 $p_{-9}/ l_{19}^{-1/2} = 0.7 \pm 0.2$
 for the low-temperature component (where $p_{\rm thm}=10^{-9} p_{-9}$ 
dyn cm$^{-2}$).
For the high-temperature component in regions 1 through 6 the 
 pressures are
 $1.4 \pm 0.4$,
but they average $5 \pm 2$
 for regions 7 through 9 because of the much higher temperatures
 determined for
 those regions, averaging 
4~keV compared to only 0.8~keV for the high-temperature
 component in regions 1$-$6 (and 0.6~keV for the low-temperature component in all regions).

\subsection{Morphology of the Ring and Cometary Nebula} \label{s:mach}

The cometary  
 shape observed from many PWNe
is  generally attributed to motion of the system relative to the ambient medium.
In the case of powerful, young, subsonically-moving pulsars such as the Crab,
 a distinct jet-torus structure may be present near the pulsar
 related to 
axial symmetry in the wind caused by rotation of the pulsar.
In other systems, particularly older pulsars moving supersonically
 through a cold interstellar medium,
 the PWN structure consists of a forward bowshock and an extended tail.
 
\cxou\ appears to have some attributes of both these extremes.
It has a distinct, nearly circular, $\sim$5\arcsec-radius circumstellar ring
 encircling jet-like features and a $\sim$2\arcmin$\times$1.5\arcmin\ comet-shaped nebula.
In analogy to the Crab,
 the ring may define a termination shock (TS) 
 where the ultrarelativistic free-flowing
equatorial
 wind slows to
 sub-relativistic speeds, referred to as the inner ring in the Crab \citep{Wei00},
 or the  ring may be a 
torus 
of shocked pulsar wind
 similar to the one
 that lies some distance beyond the inner ring of the Crab.
Since in both cases the symmetry axis of the \cxou\ ring should be the spin axis
of the pulsar, the 
ellipticity of the ring implies the angle $\zeta$  between the line of sight
and the spin axis is about $30^\circ$
(see Section~\ref{s:ring}).
If we assume that the spin axis coincides with the direction of 
motion, 
similar to many other pulsars \citep{Joh05}, and \cxou\ originated
at the center of \ic, then the pulsar velocity can be estimated as
$v_{\rm psr}\sim 400$--600 km s$^{-1}$, for $d_{1.5}=1$ and $\tau_4\sim 2$--3
(see
Section~\ref{s:centralsource}), comparable to the speed of sound in the
ambient medium estimated in Section~\ref{s:surround}.
The relatively small $\zeta$ value corresponds to a 
 distance from the SNR center to the pulsar of $s\sim 4\times 10^{19}d_{1.5}$ cm.
This requires the SNR must be elongated along the line of sight if 
 \cxou\ is still within the SNR.
For instance, if the SNR is a prolate spheroid 
 whose polar axis is parallel to the line of sight, then
 this axis must be a factor of 2 longer than the equatorial radius
and the length of the chord through the SNR along the line of sight to \cxou\
becomes $l_{19}\sim 7 d_{1.5}$ (cf. Section~\ref{s:surround}).

In this geometry the north-south jet-like features could be true pulsar
jets \citep[e.g., similar to the inner jets of the Vela pulsar;][]{Pav03}
of $\sim 10^{17} d_{1.5}$ cm (deprojected) length.
The fact that the southern jet is brighter than the northern jet can be
explained by the Doppler boosting, i.e., the outflow in the
 southern jet is approaching the observer, which implies that the pulsar
is also approaching and located in the front part of the SNR. 
The offset of the apparent ring center from the pulsar (see Figure \ref{f:ring}) could
be explained similar to that of the Crab inner ring \citep{Wei12},
e.g., as caused by an intrinsic azimuthal asymmetry due to nonuniformity
of the ambient medium or even a simple geometric displacement caused by a non-equatorial ring. 

In this interpretation the observed X-ray PWN beyond the ring is comprised
of a shocked pulsar wind outflow. Its cometary shape is formed by the ram
pressure caused by the motion of \cxou\  relative to the ambient medium.
The misalignment of $\sim 50^\circ$ between the symmetry axis of 
the comet-shaped PWN 
and the direction toward the SNR center could be explained by bulk
motion of the SNR matter in the vicinity of \cxou\ with a velocity of
a few hundred km s$^{-1}$ (e.g., due east or east-southeast).
  
We should note that the ellipticity of the \cxou\ ring derived from the
image analysis (Section 3.1.2) is subject to large systematic 
uncertainties due to the ring faintness and the (apparent) 
ring-jet intersections. Therefore, we cannot exclude the possibility that
the true ellipticity is much smaller than $30^\circ$, i.e., the ring is
nearly circular. In this case, if  we assume that the 
pulsar's spin axis is aligned with direction of motion,  
we would have to
conclude that \cxou\ could {\em not} originate at the center of the \ic\ SNR.
However, if we allow a strong misalignment between the spin axis and
the direction of motion \citep[e.g., similar to IGR\,J11014--6103; ][]{Pav14}, then 
the circular shape of the ring would not impose restrictions on the
direction of motion and birthplace of the pulsar. Nevertheless, if the jet-like
features are indeed the jets, they would lie nearly along the line
of sight and would be exceptionally long.
Thus, some ellipticity of the ring is required if 
it is associated with
the equatorial pulsar wind.

The cometary shape of the larger PWN may offer some additional guidance in
interpreting the observed properties of \cxou. 
In general, 
 the pulsar wind flows non-radiatively until it 
encounters the TS, while the shocked pulsar wind, 
 confined between the TS and a contact discontinuity (CD)
that separates the shocked pulsar wind from shock-heated ambient material,
emits synchrotron radiation.
The shapes of the TS, CD and the outer (forward) shock,
which separates the shocked ambient gas from unshocked ambient gas, depend on the
Mach number. 

In the supersonic case, ${\cal M}\gg 1$, the shapes of these surfaces
ahead of the moving pulsar resemble paraboloids
 whose sky projections look bow-like if 
the angle between the pulsar velocity and the line of sight is large
enough.
The shocked pulsar wind at the bow apex is confined in a thin layer,
 and thus the TS and CD 
surfaces upstream from the central source are close to each other,
$R_{\rm TS}(\theta) \sim R_{\rm CD}(\theta)$ at $\theta\ll 1$,
where $\theta$ is the polar angle with respect to the symmetry axis of the
flow.
The TS standoff distance  can be estimated from the balance between
 the bulk wind pressure inside the TS 
and the ram pressure of the ambient matter:
\begin{equation} \label{e:standoff}
\dot{E}/[4\pi c R_{\rm TS}^2(\theta=0)] \sim \rho v_{0}^2 
\end{equation}
 (assuming the pulsar wind is isotropic inside the TS).
At arbitrary $\theta$ the pressure balance defines a surface
 $R_{\rm TS}(\theta)$.
An analytical solution for $R_{\rm TS}(\theta)$ was
 derived 
by \citet{Wil96} in the thin-layer approximation, it 
 agrees well with numerical simulations \citep[e.g.,][]{Buc05}
 for small $\theta$ and high ${\cal M}$.
For $\theta\ll 1$, the Wilkin formula can be approximated by the parabola
 $(y/R_{0}) = 
C (x/R_{0})^2 +1$ with $C=0.3$, where $R_{0}=R_{\rm TS}(0)$, 
$-y$ is along the direction of motion, 
$(x/y) = \tan\theta$,
and the pulsar is at the focus $x=y=0$.
Although the head of the observed cometary  PWN resembles a parabola,
the fit to the X-ray contours in Section~\ref{s:bp} gives a
substantially larger
coefficient of the quadratic term, $C=0.54$ instead of 0.3.
The sharper (narrower) head of the observed PWN\footnote{Note that the parabola becomes even 
`wider' if there is a component
 of the motion along the line of sight, $i < 90\arcdeg$.}
 and the smooth surface 
brightness profile ahead of the pulsar (i.e., the lack of any
structure that could be identified with a TS; see Figure 5)  
provides additional evidence that the pulsar motion is not supersonic. 

In the transonic case, ${\cal M} \sim 1$, 
 the ambient thermal pressure contributes, in addition to the ram pressure,
 to the right hand side of Equation~(\ref{e:standoff}).
Since $c_s^2 = \gamma p_{\rm thm}/\rho$, 
 the total pressure can be written as
$p_{\rm thm} + \rho v_{0}^2 = 
 \rho v_{0}^2 
(1+\gamma^{-1}{\cal M}^{-2})$, and the stand-off distance for the CD
can be estimated as 
$R_{\rm CD}(0) \sim R_{\rm TS}(0) \sim \dot{E}^{1/2}(4\pi c \rho v_0^2)^{-1/2} 
(1+\gamma^{-1}{\cal M}^{-2})^{-1/2}
 \sim 10^{17}(\dot{E}/3\times 10^{36}\,{\rm erg\,s}^{-1})^{1/2}  
p_{-9}^{-1/2}(1+\gamma{\cal M}^2)^{-1/2}$ cm.
Inserting plausible parameter values for the pulsar and the ambient
medium (estimated in Sections~\ref{s:centralsource} and~\ref{s:surround}),
 $R_{\rm CD}(0)\sim 10^{17}$~cm which corresponds to $4.5\arcsec d_{1.5}^{-1}$ or
 roughly the observed ring radius.
In this transonic case, 
 there is no longer a general analytic solution, but the parabolic
approximation for the CD head at 
$\gamma{\cal M}^2 \gtrsim 0.7$
gives  
$C\approx 0.3+0.2(\gamma{\cal M}^2)^{-1}$. 
It becomes 
 close to the measured $C=0.54$ at $\gamma{\cal M}^2\approx 0.8$, or
${\cal M}\approx 0.7$ for $\gamma=5/3$.

In the subsonic limit, ${\cal M} \ll 1$,
 the canonical bow-shock model must be abandoned
 because the thin layer assumption, $R_{\rm TS} \sim R_{\rm CD}$, 
does not apply even in the forward direction. 
Instead, the flow can be treated as incompressible
 so that 
 Laplace's equation for the velocity field, with appropriate boundary conditions,
 correctly describes the hydrodynamics.
Near the pulsar, the flow is again idealized as purely radial,
 whereas at large distances the velocity must be uniform and directed opposite
 the direction of motion of the pulsar.
The solution is then 
the sum of 
 a uniform flow and a radial flow from the pulsar with
 speed falling as the square of the radius.
Since the post-shock radial velocity component must be about $c/3$ at the TS
outer boundary \citep{KenCor84}
 but 
$\sim v_0\ll c/3$ at the CD (where the two velocity components cancel),
 $R_{\rm TS}(0) \ll R_{\rm CD}(0)$,
 and the CD standoff distance is
$R_{\rm CD}(0) = (c/3v_0)^{1/2} R_{\rm TS}(0) \simeq 18 (v_0/300\,{\rm km\,s}^{-1})^{-1/2}
R_{\rm TS}(0)$.  
We do not clearly see the CD in our images, perhaps because it is blurred by
instabilities.
However, if we assume
$R_{\rm CD}(0)\sim 15\arcsec$--30\arcsec (see Figures 5 and 6), then
the above estimate gives $R_{\rm TS}(0)\sim 1\arcsec$--2\arcsec
(for $v_0=300$ \kms), much
smaller than the ring radius, $\sim 5\arcsec$.
This means the ring may be analogous to the Crab's torus 
while the TS may be hidden in the wings of the pulsar's PSF (see Figure \ref{f:subrp}).
Note that the relative motion cannot be highly subsonic for reasonable estimates of $v_0$
 since $c_s \sim 400$ km/s.
 
The overall shape of the PWN also depends on the Mach number.
In the supersonic and transonic regimes,
 numerical solutions \citep{Buc05} show the $R_{\rm TS}(\theta)$ surface
 develops a bullet shape
 with the TS elongated in the downstream direction ($\theta = \pi$) and
 terminating in a Mach disk oriented perpendicular to the direction of motion.
Simulations show the ratio 
 $R_{\rm TS}(\pi)/R_{\rm TS}(0) \sim (1 + {\cal M}^2)^{1/2}$
 for lower ${\cal M}$, reaching a limit of $\sim$5 for ${\cal M} \gg 1$.
This amount of elongation is incompatible with observations of \cxou, again suggesting
 the flow is not supersonic.
Note that a circular ring is only conceivable 
if the pulsar wind
 was purely radial producing a limb-brightened spherical emission region
 outside the TS.
However, pulsar winds are known to be highly aspherical resulting in
 emission morphologies like the Crab torus.
As we mentioned above, if the ring is a torus,
 then the spin axis is predominantly along the line of sight.

Finally, 
in the supersonic and transonic regimes, the expected shocked pulsar wind speeds
 in the tail (roughly along the CD) are 0.3$-$0.5$c$ 
 \citep{Buc05}, whereas they
 are only of order $v_{0} \sim 300$~\kms\ in the subsonic case.
Particles in this flow will only radiate at X-ray energies, $E \sim 1$~keV, for 
 a time $\tau \sim 200 (E/1\,{\rm keV})^{-1/2}  (B/50\,\mu{\rm G})^{-3/2}$~yr
 due to their synchrotron losses.
Here, $B$ is the equipartition magnetic field defined \citep{Pac70} as
 $B= (6 \pi c_{12} (1+k)L/(fV))^{2/7}$ 
 where $f$ is the filling factor in a volume $V$ producing a synchrotron luminosity $L$
 over some spectral range, 
 $c_{12}$ a constant dependent on this spectral range and the spectral slope (index),
 and the ion to electron energy ratio $k=0$ for a pure pair plasma.
From the flux density in the radio band, $\sim$86~mJy (section~\ref{s:r_pwn}),
 a volume of that region of $10^{54}$~cm$^3$ assuming a filled prolate ellipsoid,
 and a flat radio spectrum extending to 100~GHz ($c_{12} = 7\times 10^6$),
the field strength is $B \sim 65 (\sin i)^{2/7}$~$\mu$G.
The length of the \cxou\ nebular tail extends roughly 1.5\arcmin\
 or $2\times 10^{18} d_{1.5}(\sin i)^{-1}$~cm implying a velocity of 
 $\sim$500$(\sin i)^{-5/7}$~\kms\ for this synchrotron age.
This is somewhat larger than the subsonic velocity estimate
 and only approaches the transonic velocity estimate as $i \rightarrow 0$\arcdeg.

In summary,
 the shape of the forward region of the \cxou\ PWN and the size of its surrounding ring
 implies a transonic (or mildly subsonic) net flow
 because of the additional flow collimation that occurs at low ${\cal M}$.
The size of the circular ring is compatible with the standoff distance of a termination shock 
 unless the flow is subsonic in which case the ring is larger than the TS by a factor of 20 or so.
In any case, the near circularity of the ring implies either emission from a limb-brightened
 spherical emission region or emission from a toroidal region with symmetry axis nearly
 along the line of sight.
Neither of these two conclusions is satisfactory, however, as the former belies the known
 predominantly azimuthal emission geometry associated with rotating magnetized neutron stars
 and the latter implies a pulsar orientation unlikely to produce the observed shape of the 
 larger cometary PWN.

{\bf Acknowledgments}

The \cxo\ observations were obtained in response to Chandra Proposal Number 13500093  by the Chandra X-ray Observatory Center, which is operated by the Smithsonian Astrophysical Observatory for and on behalf of the National Aeronautics Space Administration under contract NAS8-03060.
Optical observations were obtained with the SARA Observatory 0.9~m telescope at Kitt Peak, which is owned and operated by the Southeastern Association for Research in Astronomy. 
Basic research in radio astronomy at the Naval Research Laboratory (TC) is supported by 6.1 Base funding.

\clearpage
\begin{deluxetable}{lrrrrrrrrr}
\tablecolumns{10}
\tablewidth{0pt}
\tablefontsize{\footnotesize}
\tablecaption{Outlier Regions Spectral Fit Parameters \label{t:fitparams2}}  
\tablehead{
\colhead{Region$^{\rm a}$} & \colhead{1} & \colhead{2} & \colhead{3} & \colhead{4} & \colhead{5} & \colhead{6} & \colhead{7} & \colhead{8} & \colhead{9}
 }

\startdata
Area (arcmin$^2$)      &  \multicolumn{1}{c}{4.75} & \multicolumn{1}{c}{1.96} & \multicolumn{1}{c}{1.53} & \multicolumn{1}{c}{3.80} & 
                          \multicolumn{1}{c}{2.95} & \multicolumn{1}{c}{4.23} & \multicolumn{1}{c}{5.07} & \multicolumn{1}{c}{2.02} & \multicolumn{1}{c}{3.03} \\
\cutinhead{Absorbed (Power Law $+$ Variable MeKaL)$^{\rm b}$}
$n_{\rm H}/10^{22}$ (cm$^2$)    &  0.49$^{+ 0.03}_{-0.05}$  &  0.37$^{+ 0.03}_{-0.04}$  &  0.64$^{+ 0.05}_{-0.12}$  &  0.51$^{+ 0.04}_{-0.03}$  &  0.48$^{+ 0.04}_{-0.07}$  &  0.48$^{+ 0.03}_{-0.04}$  &  0.58$^{+ 0.06}_{-0.06}$  &  0.48$^{+ 0.09}_{-0.05}$  &  0.45$^{+ 0.03}_{-0.03}$  \\
$\Gamma$                        &  2.99$^{+ 0.33}_{-0.62}$  & 4.14$^{+ 1.03}_{-0.77}$  &  3.52$^{+ 0.60}_{-0.60}$  &  3.91$^{+ 0.51}_{-0.30}$  &  3.52$^{+ 0.68}_{-1.18}$  &  3.50$^{+ 0.30}_{-0.36}$  &  2.17$^{+ 0.10}_{-0.11}$  &  1.95$^{+ 0.10}_{-0.09}$  &  2.26$^{+ 0.12}_{-0.17}$  \\
$kT$ (keV)                      &  0.69$^{+ 0.02}_{-0.02}$  &  0.70$^{+ 0.03}_{-0.01}$  &  0.79$^{+ 0.05}_{-0.16}$  &  0.69$^{+ 0.02}_{-0.02}$  &  0.68$^{+ 0.03}_{-0.02}$  &  0.66$^{+ 0.02}_{-0.02}$  &  0.68$^{+ 0.03}_{-0.02}$  &  0.78$^{+ 0.06}_{-0.06}$  &  0.68$^{+ 0.02}_{-0.02}$  \\
$K_{\rm POW}/10^{-4}$           &  1.41$^{+ 0.99}_{-0.92}$  &  0.04$^{+ 0.04}_{-0.04}$  &  1.30$^{+ 0.73}_{-0.96}$  &  2.66$^{+ 1.51}_{-0.96}$  &  0.85$^{+ 0.63}_{-0.80}$  &  1.71$^{+ 0.63}_{-0.95}$  &  3.34$^{+ 0.48}_{-0.45}$  &  2.06$^{+ 0.27}_{-0.26}$  &  2.45$^{+ 0.42}_{-0.51}$  \\
$K_{\rm THM}/10^{-4}$           & 12.10$^{+ 5.11}_{-2.48}$  & 8.20$^{+ 3.26}_{-2.19}$  &  3.26$^{+ 3.52}_{-2.28}$  &  8.34$^{+ 3.38}_{-3.71}$  &  3.01$^{+ 1.64}_{-1.16}$  &  3.57$^{+ 3.01}_{-0.23}$  &  2.88$^{+ 1.09}_{-0.35}$  &  0.87$^{+ 2.47}_{-0.15}$  &  4.16$^{+ 8.03}_{-0.51}$  \\
$f_{\rm POW}/10^{-13}$ ($^{\rm c}$) &  1.60  &  0.30  &  0.88  &  1.75  &  0.70  &  1.42  &  7.58  &  6.23  &  5.43  \\
$f_{\rm THM}/10^{-13}$ &  8.59  &  6.74  &  1.57  &  6.43  &  2.68  &  4.31  &  3.18  &  1.04  &  5.32  \\
$\chi^2$/dof                    &  356.6/238  &  251.0/180  &  176.6/161  &  346.4/222  &  314.2/188  &  359.7/220  &  507.0/278  &  307.9/259  &  381.9/262  \\
\cutinhead{Absorbed (Variable MeKaL $+$ Variable MeKaL)$^{\rm d}$}
$n_{\rm H}/10^{22}$ (cm$^2$)    &  0.47$^{+ 0.04}_{-0.05}$  &  0.33$^{+ 0.03}_{-0.03}$  &  0.53$^{+ 0.15}_{-0.10}$  &  0.47$^{+ 0.07}_{-0.05}$  &  0.45$^{+ 0.08}_{-0.09}$  &  0.48$^{+ 0.06}_{-0.06}$  &  0.56$^{+ 0.05}_{-0.06}$  &  0.49$^{+ 0.09}_{-0.08}$  &  0.46$^{+ 0.04}_{-0.04}$  \\
$kT_1$ (keV)                    &  0.60$^{+ 0.04}_{-0.04}$  &  0.65$^{+ 0.03}_{-0.03}$  &  0.50$^{+ 0.07}_{-0.07}$  &  0.60$^{+ 0.04}_{-0.04}$  &  0.56$^{+ 0.08}_{-0.08}$  &  0.55$^{+ 0.05}_{-0.04}$  &  0.64$^{+ 0.02}_{-0.02}$  &  0.64$^{+ 0.05}_{-0.05}$  &  0.67$^{+ 0.02}_{-0.02}$  \\
$kT_2$ (keV)                    &  0.78$^{+ 0.02}_{-0.02}$  &  0.74$^{+ 0.15}_{-0.02}$  &  1.02$^{+ 0.26}_{-0.09}$  &  0.80$^{+ 0.11}_{-0.05}$  &  0.77$^{+ 0.11}_{-0.05}$  &  0.82$^{+ 0.09}_{-0.05}$  &  3.55$^{+ 0.36}_{-0.30}$  &  4.72$^{+ 0.78}_{-0.50}$  &  3.75$^{+ 1.00}_{-0.51}$  \\
$K_1/10^{-4}$                   &  3.19$^{+ 3.18}_{-0.29}$  &  2.68$^{+ 6.32}_{-0.57}$  &  4.27$^{+ 2.65}_{-2.56}$  &  7.44$^{+ 6.40}_{-4.07}$  &  1.72$^{+ 1.57}_{-1.08}$  &  3.60$^{+ 3.07}_{-1.38}$  &  6.81$^{+ 1.95}_{-1.77}$  &  2.98$^{+ 1.07}_{-0.89}$  &  7.39$^{+ 4.43}_{-1.85}$  \\
$K_2/10^{-4}$                   & 16.86$^{+ 6.05}_{-9.10}$  &  9.13$^{+ 2.60}_{-7.19}$  &  6.69$^{+ 1.56}_{-2.44}$  & 12.19$^{+ 4.94}_{-7.16}$  &  6.18$^{+ 3.37}_{-3.37}$  &  6.60$^{+ 1.94}_{-3.09}$  &  7.10$^{+ 1.06}_{-1.33}$  &  5.68$^{+ 0.50}_{-1.01}$  &  4.01$^{+ 1.09}_{-0.87}$  \\
$f_1/10^{-13}$         &  2.52  &  2.39  &  0.70  &  3.37  &  0.88  &  2.32  &  3.43  &  1.22  &  5.97  \\
$f_2/10^{-13}$         &  7.41  &  4.37  &  1.70  &  4.73  &  2.43  &  3.28  &  7.09  &  5.87  &  4.79  \\
$\chi^2$/dof                    &  323.1/238  &  239.7/180  &  150.1/161  &  327.3/222  &  305.3/188  &  317.5/220  &  414.5/278  &  280.1/259  &  374.4/262  \\
\enddata
\tablenotetext{a}{Numbers correspond to the numbered regions shown in Figure~\ref{f:s3bg}}
\tablenotetext{b}{Abundances tied to solar ratios for the groups  CNO, Ne and Mg, Si and S, and Fe and Ni.}
\tablenotetext{c}{All fluxes in units of \ergcms.}
\tablenotetext{d}{Same as (b) and abundances of second MeKaL model tied to those of first model.}
\end{deluxetable}

\begin{table}[ht]
\begin{center}
\caption{Spectral Fit Parameters \& Derived Quantities for \cxoulong } \label{t:fitparams1}
\begin{tabular}{cccccccc}
\hline \hline
Model 
& \multicolumn{1}{c}{$n_{\rm H}$} 
& \multicolumn{1}{c}{$T^{\infty}$} 
& \multicolumn{1}{c}{$R^{\infty}$} 
& \multicolumn{1}{c}{$L_{\rm bol}^{\infty}$} 
& \multicolumn{1}{c}{$\Gamma$} 
& \multicolumn{1}{c}{$L_{\Gamma}$} 
& \multicolumn{1}{c}{$\chi^2/{\rm dof}$}  \\
  
& \multicolumn{1}{c}{$10^{22}$~cm$^{-2}$} 
& \multicolumn{1}{c}{eV} 
& \multicolumn{1}{c}{km} 
& \multicolumn{1}{c}{$10^{32}$ \ergl} 
& \multicolumn{1}{c}{} 
& \multicolumn{1}{c}{$10^{31}$ \ergl}
& \multicolumn{1}{c}{}  \\
\hline
NSA  & 0.63$^{+0.02}_{-0.03}$ &  58.4$^{+0.6}_{-0.4}$ & 13.05                   & 2.6$\pm$0.1 & 
    1.34$^{+0.31}_{-0.35}$ & 1.0$^{+1.4}_{-0.5}$ & 42.2/42 \\
BB   & 0.61$^{+0.13}_{-0.09}$ & 132.0$^{+8.8}_{-7.0}$ & 1.63$^{+0.09}_{-0.07}$ & 1.0$^{+0.2}_{-0.1}$ & 
    2.00$^{+0.30}_{-0.20}$ & 1.4$^{+0.6}_{-0.5}$ & 41.7/41 \\
\hline
\end{tabular}
\end{center}
\end{table}

\begin{deluxetable}{lrrrrrr}
\tablecolumns{7}
\tablewidth{0pt}
\tablefontsize{\normalsize}
\tablecaption{Upstream Spectral Fit Parameters \label{t:fitparams3}} 
\tablehead{
\colhead{Region$^a$} & \colhead{1} & \colhead{2} & \colhead{3} & \colhead{4} & \colhead{5} & \colhead{6} 
 }

\startdata
\cutinhead{Absorbed (Power Law $+$ Variable MeKaL)$^{\rm b}$}
$n_{\rm H}/10^{22}$  (cm$^2$)      &  0.61$^{+ 0.09}_{-0.08}$  &  0.40$^{+ 0.06}_{-0.04}$  &  0.40$^{+ 0.06}_{-0.05}$  &  0.40$^{+ 0.05}_{-0.05}$  &  0.40$^{+ 0.08}_{-0.06}$  &  0.37$^{+ 0.06}_{-0.05}$  \\
$\Gamma$                           &  2.08$^{+ 0.15}_{-0.15}$  &  2.13$^{+ 0.23}_{-0.24}$  &  2.19$^{+ 0.27}_{-0.28}$  &  2.59$^{+ 0.39}_{-0.38}$  &  2.41$^{+ 0.44}_{-0.47}$  &  2.43$^{+ 0.41}_{-0.42}$  \\
$kT$                  (keV)        &  0.39$^{+ 0.10}_{-0.05}$  &  0.70$^{+ 0.06}_{-0.09}$  &  0.67$^{+ 0.07}_{-0.10}$  &  0.72$^{+ 0.12}_{-0.04}$  &  0.68$^{+ 0.07}_{-0.08}$  &  0.71$^{+ 0.10}_{-0.09}$  \\
$K_{\rm POW}/10^{-5}$              &  6.50$^{+ 1.15}_{-0.98}$  &  3.01$^{+ 0.87}_{-0.71}$  &  2.45$^{+ 0.82}_{-0.66}$  &  2.39$^{+ 1.01}_{-0.78}$  &  1.59$^{+ 0.89}_{-0.66}$  &  1.62$^{+ 0.79}_{-0.61}$  \\
$K_{\rm THM}/10^{-5}$              &  7.12$^{+ 3.76}_{-2.53}$  &  3.79$^{+ 0.93}_{-0.79}$  &  2.91$^{+ 0.79}_{-0.70}$  &  2.75$^{+ 0.71}_{-0.64}$  &  3.32$^{+ 0.81}_{-0.71}$  &  2.65$^{+ 0.68}_{-0.59}$  \\
$f_{\rm POW}/10^{-13}$  (\ergcms)  &  1.60  &  0.78  &  0.60  &  0.41  &  0.32  &  0.32  \\
$f_{\rm THM}/10^{-13}$  (\ergcms)  &  0.26  &  0.31  &  0.23  &  0.22  &  0.26  &  0.23  \\
$\chi^2$/dof                       &  135.5/146  &  120.1/116  &  111.7/ 99  &   97.8/ 90  &   58.6/ 84  &   66.5/ 80  \\
\cutinhead{Absorbed (Variable MeKaL $+$ Variable MeKaL)$^{\rm c}$}
$n_{\rm H}/10^{22}$ (cm$^2$)       &  0.63$^{+ 0.10}_{-0.09}$  &  0.71$^{+ 0.10}_{-0.13}$  &  0.79$^{+ 0.14}_{-0.11}$  &  0.37$^{+ 0.06}_{-0.04}$  &  0.40$^{+ 0.06}_{-0.06}$  &  0.36$^{+ 0.07}_{-0.05}$  \\
$kT_1$               (keV)         &  0.38$^{+ 0.06}_{-0.05}$  &  0.32$^{+ 0.07}_{-0.03}$  &  0.29$^{+ 0.04}_{-0.05}$  &  0.71$^{+ 0.05}_{-0.03}$  &  0.67$^{+ 0.06}_{-0.07}$  &  0.70$^{+ 0.07}_{-0.09}$  \\
$kT_2$               (keV)         &  4.16$^{+ 1.04}_{-0.70}$  &  2.55$^{+ 0.51}_{-0.39}$  &  2.40$^{+ 0.56}_{-0.38}$  &  3.37$^{+ 1.43}_{-1.09}$  &  4.37$^{+ 8.06}_{-1.67}$  &  3.63$^{+ 3.23}_{-1.23}$  \\
$K_1/10^{-4}$                      &  0.90$^{+ 0.54}_{-0.31}$  &  1.31$^{+ 0.81}_{-0.59}$  &  1.65$^{+ 2.24}_{-0.71}$  &  0.33$^{+ 0.07}_{-0.04}$  &  0.39$^{+ 0.08}_{-0.07}$  &  0.30$^{+ 0.07}_{-0.07}$  \\
$K_2/10^{-4}$                      &  1.36$^{+ 0.11}_{-0.10}$  &  0.87$^{+ 0.08}_{-0.10}$  &  0.69$^{+ 0.08}_{-0.07}$  &  0.31$^{+ 0.08}_{-0.05}$  &  0.23$^{+ 0.07}_{-0.06}$  &  0.24$^{+ 0.07}_{-0.06}$  \\
$f_1/10^{-13}$ (\ergcms)           &  0.31  &  0.30  &  0.24  &  0.28  &  0.31  &  0.26  \\
$f_2/10^{-13}$ (\ergcms)           &  1.54  &  0.74  &  0.55  &  0.36  &  0.29  &  0.29  \\
$\chi^2$/dof                       &  144.1/146  &  134.0/116  &  115.2/ 99  &  107.9/ 90  &   65.5/ 84  &   72.3/ 80  \\
\enddata
\tablenotetext{a}{Regions upstream (southwest) of \cxou\ (see Figure~\ref{f:s3bg}), numbered by increasing distance from \cxou.}
\tablenotetext{b}{Abundances fixed to those of Outlier region 6 (see Table~\ref{t:fitparams2}.}
\tablenotetext{c}{Same as (b) and abundances of second MeKaL model tied to those of first model.}
\end{deluxetable}

\begin{deluxetable}{lrrrrrrr}
\tablecolumns{8}
\tablewidth{0pt}
\tablecaption{Pulsar Wind Nebula Spectral Fit Parameters \label{t:fitparams4}} 
\tablehead{
\colhead{Region$^{\rm a}$} & \colhead{1} & \colhead{2} & \colhead{3} & \colhead{4} & \colhead{5} & \colhead{6} & \colhead{7} \\
 }

\startdata
Area (arcmin$^2$)     &  \multicolumn{1}{c}{0.048} & \multicolumn{1}{c}{0.091} & \multicolumn{1}{c}{0.157} & \multicolumn{1}{c}{0.266} & 
                         \multicolumn{1}{c}{0.365} & \multicolumn{1}{c}{0.531} & \multicolumn{1}{c}{0.722} \\
\cutinhead{Absorbed Power Law}
$n_{\rm H}/10^{22}$  (cm$^2$)    &  0.56$^{+ 0.04}_{-0.04}$  &  0.53$^{+ 0.03}_{-0.03}$  &  0.56$^{+ 0.04}_{-0.03}$  &  0.52$^{+ 0.03}_{-0.03}$  &  0.51$^{+ 0.03}_{-0.03}$  &  0.47$^{+ 0.03}_{-0.03}$  &  0.41$^{+ 0.02}_{-0.02}$  \\
$\Gamma$                         &  1.50$^{+ 0.06}_{-0.03}$  &  1.54$^{+ 0.06}_{-0.05}$  &  1.69$^{+ 0.06}_{-0.03}$  &  1.78$^{+ 0.06}_{-0.06}$  &  1.91$^{+ 0.06}_{-0.06}$  &  2.06$^{+ 0.06}_{-0.06}$  &  2.28$^{+ 0.07}_{-0.07}$  \\
$K_{\rm POW}/10^{-4}$            &  1.35$^{+ 0.10}_{-0.09}$  &  1.54$^{+ 0.11}_{-0.10}$  &  1.79$^{+ 0.13}_{-0.11}$  &  1.84$^{+ 0.12}_{-0.12}$  &  1.98$^{+ 0.14}_{-0.13}$  &  2.02$^{+ 0.14}_{-0.13}$  &  2.08$^{+ 0.14}_{-0.13}$  \\
$f_{\rm POW}/10^{-13}$  (\ergcms)&  6.84  &  7.56  &  7.20  &  6.72  &  6.20  &  5.42  &  4.61  \\
$f^{\rm cor}_{\rm POW}/10^{-13}$ (\ergcms)  &  9.14  & 10.11  & 10.21  &  9.71  &  9.40  &  8.61  &  7.78  \\
$\chi^2$/dof                     &  289.2/293  &  336.7/302  &  278.9/295  &  326.5/285  &  294.4/282  &  353.2/274  &  375.6/254  \\
\cutinhead{Absorbed (Power Law $+$ Variable MeKaL)$^{\rm b}$}
$n_{\rm H}/10^{22}$  (cm$^2$)    &  0.64$^{+ 0.35}_{-0.07}$  &  0.67$^{+ 0.25}_{-0.16}$  &  0.75$^{+ 0.28}_{-0.11}$  &  0.77$^{+ 0.18}_{-0.15}$  &  0.67$^{+ 0.18}_{-0.10}$  &  0.60$^{+ 0.15}_{-0.09}$  &  0.47$^{+ 0.09}_{-0.07}$  \\
$\Gamma$                         &  1.55$^{+ 0.09}_{-0.04}$  &  1.53$^{+ 0.03}_{-0.03}$  &  1.77$^{+ 0.14}_{-0.13}$  &  1.89$^{+ 0.05}_{-0.11}$  &  1.95$^{+ 0.12}_{-0.09}$  &  2.03$^{+ 0.12}_{-0.09}$  &  1.97$^{+ 0.10}_{-0.09}$  \\
$kT$   (keV)                     &  0.11$^{+ 0.15}_{-0.03}$  & 0.36$^{+0.06}_{-0.28}$  &  0.36$^{+ 5.12}_{-0.10}$  &  0.32$^{+ 0.09}_{-0.04}$  &  0.29$^{+ 0.04}_{-0.05}$  &  0.37$^{+ 0.03}_{-0.06}$  &  0.67$^{+ 0.11}_{-0.06}$  \\
$K_{\rm POW}/10^{-4}$            &  1.46$^{+ 0.31}_{-0.13}$  &  1.49$^{+ 0.47}_{-0.47}$  &  2.05$^{+ 0.45}_{-0.37}$  &  2.20$^{+ 0.36}_{-0.33}$  &  2.14$^{+ 0.36}_{-0.71}$  &  1.98$^{+ 0.33}_{-0.23}$  &  1.41$^{+ 0.47}_{-0.16}$  \\
$K_{\rm THM}/10^{-4}$            & 69.94$^{+3837.17}_{-69.89}$  &  1.02$^{+ 6.70}_{-1.02}$  &  0.35$^{+ 4.52}_{-0.27}$  &  0.79$^{+ 7.59}_{-0.48}$  &  1.89$^{+ 8.93}_{-1.37}$  &  0.69$^{+ 3.03}_{-0.24}$  &  0.61$^{+ 1.14}_{-0.12}$  \\
$f_{\rm POW}/10^{-13}$ (\ergcms) &  6.81  &  7.56  &  6.99  &  6.44  &  5.98  &  5.20  &  4.10  \\
$f_{\rm THM}/10^{-13}$ (\ergcms) &  0.00  &  0.00  &  0.16  &  0.24  &  0.21  &  0.28  &  0.70  \\
$f^{\rm cor}_{\rm POW}/10^{-13}$ (\ergcms)&  9.40  &  10.10  & 10.85  & 10.60  &  9.81  &  8.62  &  6.39  \\
$f^{\rm cor}_{\rm THM}/10^{-13}$ (\ergcms)&  0.80  &  0.08  &  2.03  &  3.57  &  2.46  &  1.62  &  1.98  \\
$\chi^2$/dof                     &  284.8/287  &  336.7/301$^{\rm c}$  &  267.2/289  &  298.0/279  &  258.8/276  &  276.3/268  &  215.5/248  \\
\cutinhead{Absorbed (Variable MeKaL $+$ Variable MeKaL)$^{\rm d}$}
$n_{\rm H}/10^{22}$  (cm$^2$)    &  0.54$^{+ 0.07}_{-0.05}$  &  0.53$^{+ 0.07}_{-0.06}$  &  0.65$^{+ 0.18}_{-0.14}$  &  0.51$^{+ 0.15}_{-0.05}$  &  0.69$^{+ 0.08}_{-0.15}$  &  0.44$^{+ 0.10}_{-0.04}$  &  0.41$^{+ 0.07}_{-0.06}$  \\
$kT_1$  (keV)                    &  0.08$^{+79.82}_{-0.00}$  &  1.03$^{+78.87}_{-0.79}$  &  0.63$^{+ 0.19}_{-0.24}$  &  0.46$^{+ 0.16}_{-0.13}$  &  0.71$^{+ 0.07}_{-0.05}$  &  0.41$^{+ 0.35}_{-0.04}$  &  0.65$^{+ 0.06}_{-0.06}$  \\
$kT_2$     (keV)                 & 14.70$^{+ 4.27}_{-3.12}$  & 23.60$^{+56.30}_{-19.58}$  &  9.11$^{+ 2.85}_{-1.61}$  &  7.86$^{+ 1.65}_{-1.37}$  &  7.98$^{+ 2.20}_{-1.80}$  &  4.98$^{+ 1.18}_{-0.58}$  &  5.37$^{+ 1.23}_{-0.81}$  \\
$K_1/10^{-4}$                    & 305.73$^{+342.30}_{-305.73}$  &  1.67$^{+ 2.49}_{-1.67}$  &  1.59$^{+ 2.67}_{-1.27}$  &  1.83$^{+ 5.53}_{-1.32}$  &  3.35$^{+ 1.72}_{-1.66}$  &  2.22$^{+ 3.76}_{-1.16}$  &  2.69$^{+ 1.85}_{-1.01}$  \\
$K_2/10^{-4}$                    &  5.73$^{+ 0.40}_{-0.95}$  &  5.95$^{+ 0.34}_{-5.95}$  &  5.67$^{+ 0.72}_{-1.61}$  &  5.89$^{+ 0.44}_{-0.36}$  &  4.44$^{+ 0.92}_{-0.87}$  &  5.25$^{+ 0.37}_{-0.52}$  &  3.81$^{+ 0.35}_{-0.40}$  \\
$f_1/10^{-13}$   (\ergcms)       &  0.38  &  0.91  &  1.70  &  0.75  &  4.25  &  1.02  &  2.18  \\
$f_2/10^{-13}$   (\ergcms)       &  6.35  &  6.62  &  5.41  &  5.84  &  1.97  &  4.30  &  2.54  \\
$f^{\rm cor}_{1}/10^{-13}$ (\ergcms) &  0.38  &  0.91  &  1.70  &  0.75  &  4.25  &  1.02  &  2.18  \\
$f^{\rm cor}_{2}/10^{-13}$ (\ergcms) &  8.74  &  9.11  &  9.44  &  8.63  &  7.65  &  6.96  &  5.32  \\
$\chi^2$/dof                     &  280.0/287  &  337.3/296  &  266.0/289  &  301.4/279  &  254.2/276  &  277.5/268  &  212.4/248  \\
\enddata
\tablenotetext{a}{Regions ordered outward from \cxou\ numbering 1 through 7, respectively.}
\tablenotetext{b}{Abundances tied to solar ratios for the groups  CNO, Ne and Mg, Si and S, and Fe and Ni.}
\tablenotetext{c}{Abundances frozen to best-fit values before errors on free parameters estimated.}
\tablenotetext{d}{Same as (b) and abundances of second MeKaL model tied to those of first model.}
\end{deluxetable}

\begin{deluxetable}{lrrrrr}
\tablecolumns{6}
\tablewidth{0pt}
\tablecaption{Pulsar Wind Nebula X-Ray \& Radio Spectral Fit Parameters \label{t:Ellipsesfitpars}}
\tablehead{
\colhead{Region$^a$} & \colhead{1} & \colhead{2} & \colhead{3} & \colhead{4} & \colhead{5}  
 }
\startdata
Area (arcmin$^2$)               & 0.05        & 0.10         & 0.14         & 0.22         & 0.34       \\  \hline 
$n_{\rm H}/10^{22}$ (cm$^2$)       &  0.40$^{+ 0.03}_{-0.03}$  &  0.53$^{+ 0.04}_{-0.03}$  &  0.57$^{+ 0.03}_{-0.03}$  &  0.52$^{+ 0.03}_{-0.03}$  &  0.51$^{+ 0.03}_{-0.03}$  \\
$\Gamma$                           &  1.47$^{+ 0.05}_{-0.05}$  &  1.61$^{+ 0.06}_{-0.06}$  &  1.73$^{+ 0.06}_{-0.06}$  &  1.81$^{+ 0.06}_{-0.06}$  &  1.93$^{+ 0.06}_{-0.06}$  \\
$K_{\rm POW}/10^{-4}$              &  1.38$^{+ 0.08}_{-0.08}$  &  1.58$^{+ 0.12}_{-0.11}$  &  1.77$^{+ 0.13}_{-0.12}$  &  1.64$^{+ 0.12}_{-0.11}$  &  1.88$^{+ 0.13}_{-0.12}$  \\
$f_{\rm POW}/10^{-13}$  (\ergcms)  &  7.67  &  7.02  &  6.74  &  5.76  &  5.73  \\
$f^{cor}_{\rm POW}/10^{-13}$ (\ergcms)&  9.65  &  9.61  &  9.73  &  8.42  &  8.76  \\
$\chi^2$/DOF                       &  299.6/305  &  271.8/289  &  274.1/285  &  285.5/269  &  370.3/268  \\
$f_{330 \rm MHz}$ (mJy)               & 5.0$\pm$0.9 & 11.0$\pm$3.1 & 11.5$\pm$5.6 & 10.5$\pm$8.4 & 18.0$\pm$13.0 \\
$\tilde{\nu}$ (GHz)                   &  0.02$^{+0.16}_{-0.02}$ & 1.12$^{+5.1}_{-1.0}$ & 27$^{+66}_{-20}$ & 128$^{+218}_{-88}$ & 826$^{+945}_{-479}$ \\
\enddata
\tablenotetext{a}{Elliptical annuli numbered from inner to outer 1 to 5} 
\end{deluxetable}

\end{document}